\documentclass[10pt,amsmath,amssymb,pra,twocolumn]{revtex4-1}
\usepackage{graphicx}
\usepackage{dcolumn}
\usepackage{bm}
\usepackage{booktabs}
\usepackage{color}
\newcounter{one}
\newcommand{\bra}[1]{\langle #1 |}
\newcommand{\ket}[1]{| #1 \rangle}

\newcommand{\Tr}[0]{ \textrm{Tr}}

\newtheorem{stheorem}{Summpelemtary Theorem}

\def\QED{\mbox{\rule[0pt]{1.5ex}{1.5ex}}}

\def\endproof{\hspace*{\fill}~\QED\par\endtrivlist\unskip}

\newenvironment{proofof}[1]{\vspace*{5mm} \par \noindent
         {\bf Proof of #1:\hspace{2mm}}}{\endproof
}

\newcommand\calC{{\cal C}}
\newcommand\calD{{\cal D}}

\newcommand\calL{{\cal L}}

\newcommand\calP{{\cal P}}

\newcommand{\cl}{\mathrm{cl}}
\newcommand{\qm}{\mathrm{qm}}

\newcommand{\bd}{\mathrm{bd}}
\newcommand{\sd}{\mathrm{sd}}

\newcommand{\beq}{\begin{equation}}
\newcommand{\eeq}{\end{equation}}

\newcommand{\affA}{
Graduate School of Informatics and Engineering,
The University of Electro-Communications,
1-5-1 Chofugaoka, Chofu, Tokyo 182-8585, Japan
}
\newcommand{\affB}{Theoretical Physics Laboratory, RIKEN Cluster for Pioneering Reserach, Wako-shi, Saitama 351-0198, Japan}

\begin{document}
\title{Superconducting-like heat current:\\
Effective cancellation of current-dissipation trade off by quantum coherence}
\author{Hiroyasu Tajima}
\affiliation{\affA}
\author{Ken Funo}
\affiliation{\affB}

 
\maketitle

\textbf{
Producing a large current typically requires large dissipation, as is the case in electric conduction, where Joule heating is proportional to the square of the current. 
Stochastic thermodynamics~\cite{Seifert,Sekimoto} offers a framework to study nonequilibrium thermodynamics of small fluctuating systems, and quite recently, microscopic derivations and universal understanding of the trade-off relation between the current and dissipation have been put forward~\cite{TUreview,TUreview2,TU,SST,ST,SS,ID,AD}.  
Here we establish a universal framework clarifying how quantum coherence affects the trade-off between the current and dissipation: 
a proper use of coherence enhances the heat current without increasing dissipation, i.e. coherence can reduce friction. If the amount of coherence is large enough, this friction becomes virtually zero, realizing a superconducting-like ``dissipation-less'' heat current. 
Since our framework clarifies a general relation among coherence, energy flow, and dissipation, it can be applied to many branches of science.
As an application to energy science, we construct a quantum heat engine cycle that exceeds the power-efficiency bound on classical engines~\cite{SST}, and effectively attains the Carnot efficiency with finite power in fast cycles.
We discuss important implications of our findings with regard to the field of quantum information theory, condensed matter physics and biology.
}

Speeding-up physical processes inevitably induces ``friction''. This intuition has been well understood in the context of finite-time quantum operation~\cite{QSL,Funo-QSL,STA1,STA2} and thermodynamic control~\cite{TUreview,TUreview2,TU,SST,ST,SS,ID,AD}. 
In particular, the thermodynamic uncertainty relation \cite{TUreview,TUreview2,TU,AD} shows that the precision of the  current is constrained by dissipation, and the power-efficiency trade-off relation \cite{SST,ST,SS} in heat engines shows that producing a large output power induces dissipation, lowering the heat-to-work conversion efficiency. These trade-off relations between current and dissipation have attracted considerable attention in the field of stochastic thermodynamics, since they give universal constraints on thermodynamic quantities in a finite-time, out of equilibrium settings. 

Although significant progress has been made in recent years to understand the fundamental limits set by thermodynamics, it is still unclear, and even controversial, about the effect of quantum coherence in thermodynamics~\cite{delCampo,Brandner,Petruccione,Coherence1,Coherence2}.
This should be contrasted with other fields, such as in quantum cryptography and quantum error correction, where a proper use of quantum coherence protects the reversibility of the system~\cite{Nielsen}. We therefore expect that quantum coherence can be utilized to reduce ``friction'' in thermodynamics.


In this paper, we construct a general theoretical framework clarifying how quantum coherence affects the current-dissipation trade-off relation, and show that the above expectation is true.
Our main results indicate that quantum coherence can enhance the heat current without increasing dissipation.
In particular, we show an interesting scaling behavior that for a large amount of coherence, the heat current scales as a macroscopic order while keeping dissipation at a constant order, realizing a ``dissipation-less'' current.

Our framework provides a general classification on the types of quantum coherence that induce gains or losses in the thermodynamic performance. We find that coherence between energy eigenstates with different energies always induces losses. This is consistent with previous observations that coherence between the ground and excited states that is built up during a heat-engine cycle degrades its performance, sometimes termed as the effect of ``quantum friction''~\cite{delCampo,Brandner}. On the other hand, we find that coherence among degenerate energy eigenstates leads to gains, working as ``quantum lubrication''.
It is interesting to point out that the above classification is directly related to two important types of quantum coherence in quantum information theory, that is, the speakable and unspeakable coherence \cite{Marvian}. As a result, our framework gives thermodynamic meanings to the classification of coherence. 

Since our framework provides a unified understanding among thermodynamic irreversibility, the energy flow and quantum coherence, it has many applications in physics. 
As an application to energy science, we consider a quantum heat engine that utilizes quantum coherence. 
We give a general condition about which type of quantum coherence enhances the power and efficiency of heat engines, and construct several examples that exceed a universal power-efficiency trade-off relation~\cite{SST} for classical engines. 
In addition, we show that the ``dissipation-less'' current-driven quantum heat engine approximately attains the Carnot efficiency with finite power in fast cycles.
In view of recent proposals on the equivalence between quantum heat engines and natural and artificial light-harvesting systems~\cite{Scully,Chin}, we also discuss possible directions of using our results to understand the role of coherence and its impact on the energy transfer efficiency in light-harvesting systems. 

We consider a 
system connected to a heat bath whose inverse temperature is $\beta$. 
We assume that the time evolution of the reduced density matrix of the system $\rho$ obeys the standard quantum master equation~\cite{Breuer}:
\begin{align}
\frac{\partial \rho}{\partial t}&=-\frac{i}{\hbar}[H,\rho]+\sum_{\omega}\gamma(\omega)\left[L_{\omega}\rho L^{\dagger}_{\omega}-\frac{1}{2}\{L^{\dagger}_{\omega}L_{\omega},\rho\}\right] . \label{master}
\end{align}
Here $H$ is the Hamiltonian of the system which may have degeneracy, and the Lindblad operator $L_{\omega}$ describes a quantum jump between energy eigenstates with energy difference being $\hbar\omega$: $[L_{\omega},H]=\hbar\omega L_{\omega}$~\cite{Breuer}. 
The positive coefficient $\gamma(\omega)$ is assumed to satisfy the detailed balance relation $\gamma(\omega)/\gamma(-\omega)=\exp[\beta\hbar\omega]$. 
Note that an extension of our formulation to the case with multiple heat baths is straightforward. 

Our purpose is to clarify how coherence affects the trade-off relation between the energy flow and dissipation.
For this purpose, we focus on the ratio between the heat current $J$ and the entropy production rate $\dot{\sigma}$: $J^2/\dot{\sigma}$.
Here $J(\rho):=\Tr[H\partial_t\rho]$ is the heat current which describes the energy flow from the heat bath to the system~\cite{Funo-review}.
Also, the entropy production rate is defined as $\dot{\sigma}(\rho):=\dot{S}(\rho)-\beta J(\rho)\geq 0$, which is a key quantity that measures dissipation (thermodynamic irreversibility) in stochastic thermodynamics~\cite{Funo-review}.
Here, $\dot{S}(\rho)=\Tr[\partial_{t} \rho \log \rho]$ is the von Neumann entropy flux of the system~\cite{Nielsen}, and $\beta J$ is interpreted as the entropy increase in the heat bath. Therefore, the entropy production quantifies the total amount of entropy that is produced in the entire system, and the second law of thermodynamics is obtained as a direct consequence of the nonnegativity of $\dot{\sigma}$. 
We point out that the ratio $J^2/\dot{\sigma}$ can be used as an indicator for the performance of the heat engines, because loosely speaking, large $J$ corresponds to a large output power and small $\dot{\sigma}$ corresponds to a large heat-to-work conversion efficiency. 


To evaluate the effect of coherence on the ratio $J^2/\dot{\sigma}$, we denote the energy eigenstates of the Hamiltonian as $|e,j\rangle$, where $e$ is the energy eigenvalue and $j$ is introduced to label degenerate states. We introduce two diagonalized states $\rho_{\bd}:=\sum_{e}\Pi_e\rho\Pi_e$ and $\rho_{\sd}:=\sum_{e,j}\Pi_{e,j}\rho\Pi_{e,j}$, where $\Pi_{e,j}=|e,j\rangle\langle e,j|$ and $\Pi_{e}=\sum_{j}\Pi_{e,j}$ is the projection to the eigenspace of $H$ whose eigenvalue is $e$. 
The subscript `bd' and `sd' are the abbreviations of `block-diagonalized' and `strictly-diagonalized'.
In the state $\rho_{\bd}$, coherence among degenerate energy eigenstates is kept, but coherence among different energy eigenspaces is lost.
In the state $\rho_{\sd}$, on the other hand, all coherence is lost.
Note that if the Hamiltonian is non-degenerate, $\rho_{\bd}=\rho_{\sd}$.

Now, let us discuss how the quantum coherence affects the current-dissipation ratio $J^{2}/\dot{\sigma}$. 
We first show that coherence among different eigenspaces \textit{does not} enhance the current-dissipation ratio:
\begin{align}
\frac{J^2(\rho)}{\dot{\sigma}(\rho)}\le\frac{J^2(\rho_{\bd})}{\dot{\sigma}(\rho_{\bd})}\label{no-go}
\end{align}
The derivation of this inequality is given in Supplementary Information~\ref{go-and-no-go}.
According to~\eqref{no-go}, coherence among different eigenspaces only decrease the ratio $J^2/\dot{\sigma}$.
Namely, if the system Hamiltonian has no degeneracy, quantum coherence does not improve the performance of heat engines.


We next show that coherence among degenerate energy eigenstates \textit{does} enhance the current-dissipation ratio:
\begin{align}
\frac{J^2(\rho_{\sd})}{\dot{\sigma}(\rho_{\sd})} &\le \frac{A_{\cl}}{2} ,\label{gosd} \\
\frac{J^2(\rho_{\bd})}{\dot{\sigma}(\rho_{\bd})} &\le \frac{A_{\cl}+A_{\qm}}{2}, \label{go}
\end{align}
where the quantities $A_{\cl}$ and $A_{\qm}$ are non-negative real numbers, given by $A_{\cl}:=\Tr[X\rho_{\sd}]$ and $A_{\qm}:=\calC_{X}\calC_{l_1}(\rho_{\bd})$, with  $X:=\sum_{\omega}(\hbar\omega)^{2}\gamma(\omega) L^{\dagger}_{\omega}L_{\omega}$ and $\calC_{X}:=\max_{e, j, j': j\ne j'}|\bra{e,j}X\ket{e,j'}|$.
The derivations of~\eqref{gosd} and~\eqref{go} are shown in Supplementary Information~\ref{go-and-no-go}.
The quantity $\calC_{l_1}(\rho_{\bd})$ is the coherence $l_1$-norm with respect to the eigenstates of the Hamiltonian, which is the summation of the absolute value of the non-diagonal elements: $\calC_{l_1}(...):=\sum_{(e,j)\ne(e',j')}|\bra{e,j}...\ket{e',j'}|$.
The coherence $l_1$-norm is a well-known coherence measure in the resource theory of coherence \cite{L1norm}, and thus $A_{\rm qm}$ depends on the amount of coherence among degenerate energy eigenstates. 

Inequalities \eqref{go} and \eqref{gosd} provide general upper bounds on the current-dissipation ratio $J^2/\dot{\sigma}$ with and without coherence, respectively.
When the state has no coherence, inequality \eqref{gosd} gives a ``classical'' upper bound $A_{\cl}/2$ on the current-dissipation ratio.  
Namely, heat engines without coherence (i.e. classical heat engines) never exceed this bound.
On the other hand, inequality \eqref{go} implies that coherence among degenerate eigenstates allows the current-dissipation ratio to exceed its classical limit, up to $A_{\qm}/2$. 
We note that by combining~\eqref{no-go} and~\eqref{go}, the upper bound $(A_{\rm cl}+A_{\rm qm})/2$ also applies to a general state $\rho$, and there exists a competition between the coherence among energy eigenspaces induced losses~\eqref{no-go} and coherence among degenerate eigenstates induced gains~\eqref{go} for the current-dissipation ratio. 
Later, we give a quantum heat engine example that demonstrates this coherence-induced gains
, and show that it can operate beyond the universal limitation set on classical heat engines. 

We further find an interesting scaling behavior in~\eqref{go} as follows. 
Suppose that $A_{\qm}$ be $O(N^2)$, where $2N$ is the number of degeneracy in the system Hamiltonian. 
Then, the upper bound of the ratio $J(\rho)^2/\dot{\sigma}(\rho)$ becomes $O(N^2)$, which allows an $O(1)$ entropy production rate with an $O(N)$ heat current. 
In other words, our inequality \eqref{go} implies that large non-diagonal elements might cause macroscopic current without macroscopic dissipation.

\begin{figure}[t]
\begin{center}
\includegraphics[width=.45\textwidth]{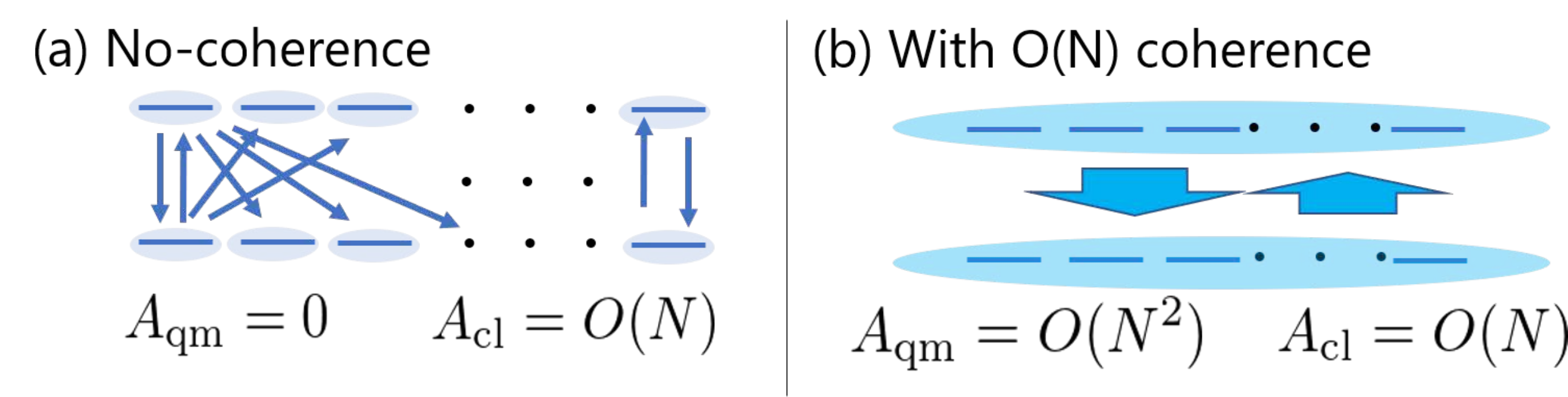}
\caption{Schematic diagram of the $2N$-state model.
(a) No coherence ($\rho=\rho_{\sd}$). In this case, correlated decays and excitations do not occur, and $A_{\cl}=O(N)$ and $A_{\qm}=0$ hold for arbitrary $\rho$. Therefore, in order to obtain $O(N)$ heat current, dissipation inevitably scales as $O(N)$.
(b) With $O(N)$ coherence  (e.g. $\rho^{+}$).
In this case, correlated decays and excitations occur and $A_{\cl}=O(N)$ and $A_{\qm}=O(N^2)$ hold.
As a result, an $O(N)$ heat current with a constant-order dissipation is realized.
}
\label{fig-2Nstate}
\end{center}
\end{figure}

The above type of current without dissipation can be realized in a concrete model
using the $2N$-state Hamiltonian, given by
\begin{align}
H=\sum^{N}_{j=1}\hbar\omega_{0}\ket{{\rm e},j}\bra{{\rm e},j}, \label{2NHamiltonian}
\end{align}
where $\ket{{\rm g},j}$ and $\ket{{\rm e},j}$ are the $j$-th degenerate ground state and excited state, respectively, and $\hbar\omega_{0}$ is the energy gap (see also Fig.~\ref{fig-2Nstate}). 
The Lindblad operators in Eq.~\eqref{master} are given by $L_{\omega_{0}}=\sum_{j,j'} \ket{{\rm e},j}\bra{{\rm g},j'}$ and $L_{-\omega_{0}}=\sum_{j,j'}\ket{{\rm g},j}\bra{{\rm e},j'}$, describing correlated decays and excitations, respectively.
Now, let us consider the state $\rho^{+}:=p_{\rm g}\ket{{\rm g},+}\bra{{\rm g},+}+p_{\rm e}\ket{{\rm e},+}\bra{{\rm e},+}$, which has large amount of coherence: $\calC_{l_1}(\rho^{+})=\calC_{l_1}(\rho^{+}_{\bd})=O(N)$. Here, $p_{\rm g}/p_{\rm e}=(1+1/N)e^{\beta\hbar\omega_{0}}$, $\ket{{\rm g},+}:=\sum_{j}\ket{{\rm g},j}/\sqrt{N}$, and $\ket{{\rm e},+}:=\sum_{j}\ket{{\rm e},j}/\sqrt{N}$.
As a result, $A_{\qm}=O(N^2)$ and the heat current becomes $O(N)$ while keeping the entropy production rate at $O(1)$, realizing a ``dissipation-less'' current (see Supplementary Information~\ref{N+1} for details):
\begin{align}
J(\rho^{+})&=N\hbar\omega_{0} \gamma(\omega_{0})p_{\rm e}=O(N),\\
\dot{\sigma}(\rho^{+})&=N\log\left(1+\frac{1}{N}\right)\gamma(\omega_{0})p_{\rm e}=O(1).
\end{align}
Moreover, we can easily generalize our example
and produce a steady-state current without dissipation by attaching the system to two heat baths (see Supplementary Information~\ref{N+1}).  

We emphasize that in the above $2N$-state model, the dissipation-less current cannot occur without quantum coherence as discussed below (see also Fig.~\ref{fig-2Nstate}). 
As we show in Supplementary Information~\ref{2N-A}, any $\rho_{\sd}$ leads to $A_{\cl}=O(N)$.
Therefore, when there is no coherence, i.e., $\rho=\rho_{\sd}$, inequality \eqref{gosd} implies that producing an $O(N)$ current is possible only when the entropy production rate is at least $O(N)$. 
This fact shows that quantum coherence causes a qualitative change in the current-dissipation trade-off relation.
Without coherence, $J$ and $\dot{\sigma}$ scales at the same order [$O(N)$]. However, with $O(N)$ coherence, $\dot{\sigma}$ can be suppressed to $O(1)$, while $J$ stays at macroscopic order [$O(N)$], realizing a dissipation-less current. 

\begin{figure}[b]
\begin{center}
\includegraphics[width=.5\textwidth]{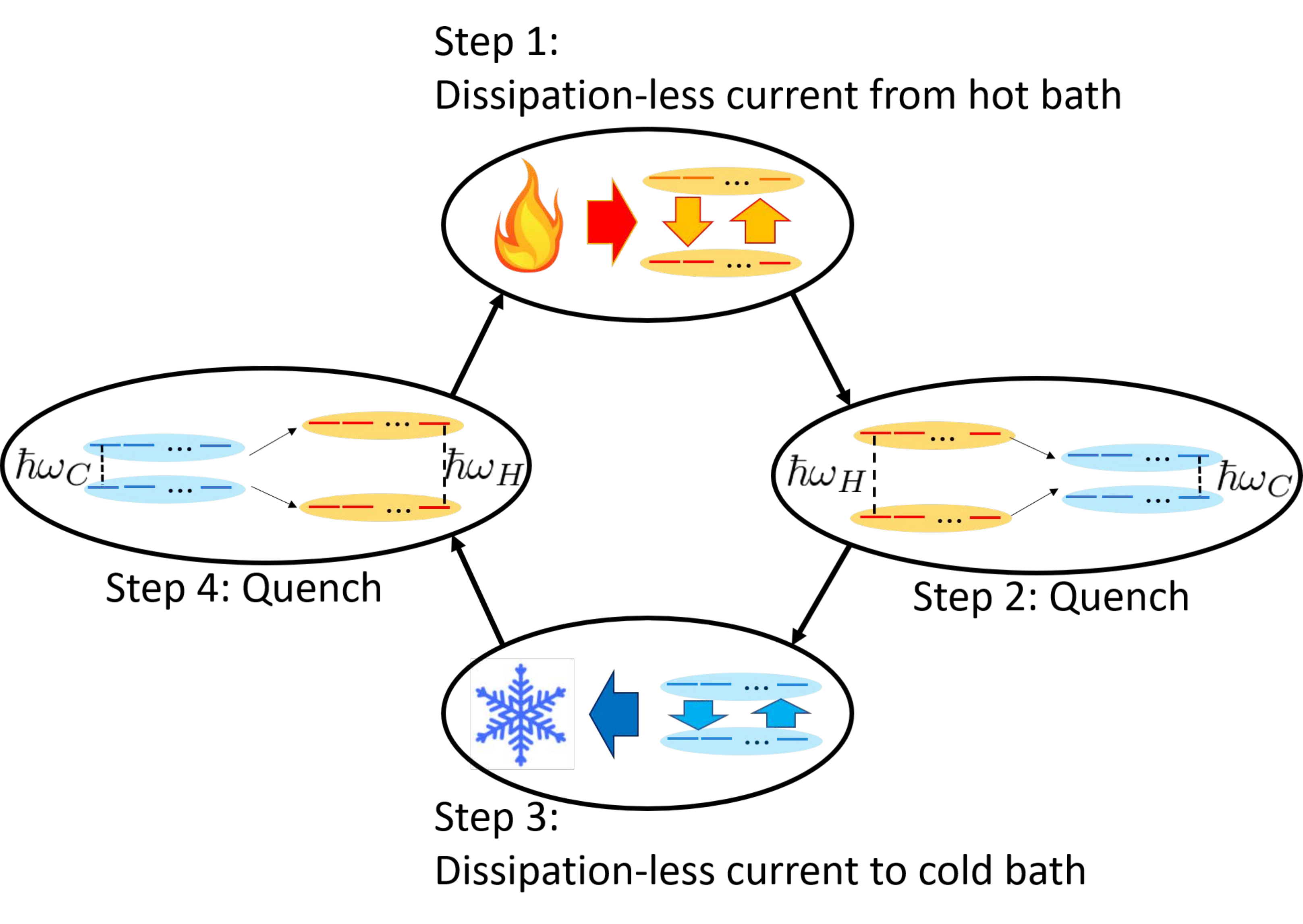}
\caption{Schematic diagram of the fast cycle attaining Carnot efficiency with finite power.
The cycle consists of four steps.
Step 1: the $2N$-state system is connected to the hot bath, absorbing the ``dissipation-less'' heat.
Step 2: the interaction between the system and the hot bath is turned off, and the energy gap of the system is changed from $\hbar\omega_H$ to $\hbar\omega_C$.
Step 3: the system is connected to the cold bath, releasing the ``dissipation-less'' heat.
Step 4: the interaction between the system and the cold bath is turned off, and the energy gap of the system is changed from $\hbar\omega_C$ to $\hbar\omega_H$. 
In this cycle, the output power scales as $O(N)$ while the thermodynamic efficiency asymptotically reaches the Carnot efficiency: $\eta=\eta_{\rm Car}-O(1/N)$. 
}
\label{fig-Carnot}
\end{center}
\end{figure}

By utilizing the dissipation-less current that appears in the $2N$-state model, we can construct a fast heat engine cycle which approximately attains the Carnot efficiency with finite output power.
In what follows, we briefly explain each step of the heat engine cycle (see also Fig.~\ref{fig-Carnot}).
1. We turn on the interaction between the system and the hot heat bath, whose inverse temperature is $\beta_{\rm H}$. The system absorbs the ``dissipation-less'' heat $Q_{\rm H}=\int J_{\rm H}dt>0$ from the hot bath, with a time-duration $\tau_{\rm H}$.
2. We turn off the interaction between the system and the hot bath, and change the energy gap from $\hbar\omega_{\rm H}$ to $\hbar\omega_{\rm C}$. 
3. We turn on the interaction between the system and the cold heat bath, whose inverse temperature is $\beta_{\rm C}$. The system releases the ``dissipation-less'' heat $Q_{\rm C}=-\int J_{\rm C}dt>0$ to the cold bath, with a time-duration $\tau_{\rm C}$.
4. We turn off the interaction between the system and the cold bath, and restore the energy gap to its initial one ($\hbar\omega_{\rm H}$).
For a stationary cycle (i.e. a cycle whose initial and final states are the same), the first law of thermodynamics implies that the extracted work is given by $W=Q_{\rm H}-Q_{\rm C}$. The output power is then defined as the work per unit time: $W/\tau$, where $\tau=\tau_{\rm H}+\tau_{\rm C}$. The thermodynamic efficiency is defined as $\eta=W/Q_{\rm H}$, which quantifies the heat-to-work conversion ratio. Note that $\eta$ is always bounded from above by the Carnot efficiency $\eta_{\rm Car}=\beta_{\rm H}/\beta_{\rm C}$, as a direct consequence of the second law.   
As we discuss in Supplementary Information~\ref{N+1}, the cycle time of our heat engine can be shorter than the typical relaxation time of the system, and the output power scales as $O(N)$ while the thermodynamic efficiency  asymptotically reaches the Carnot efficiency: $\eta=\eta_{\rm Car}-O(1/N)$. 

\begin{figure}[t]
\begin{center}
\includegraphics[width=.4\textwidth]{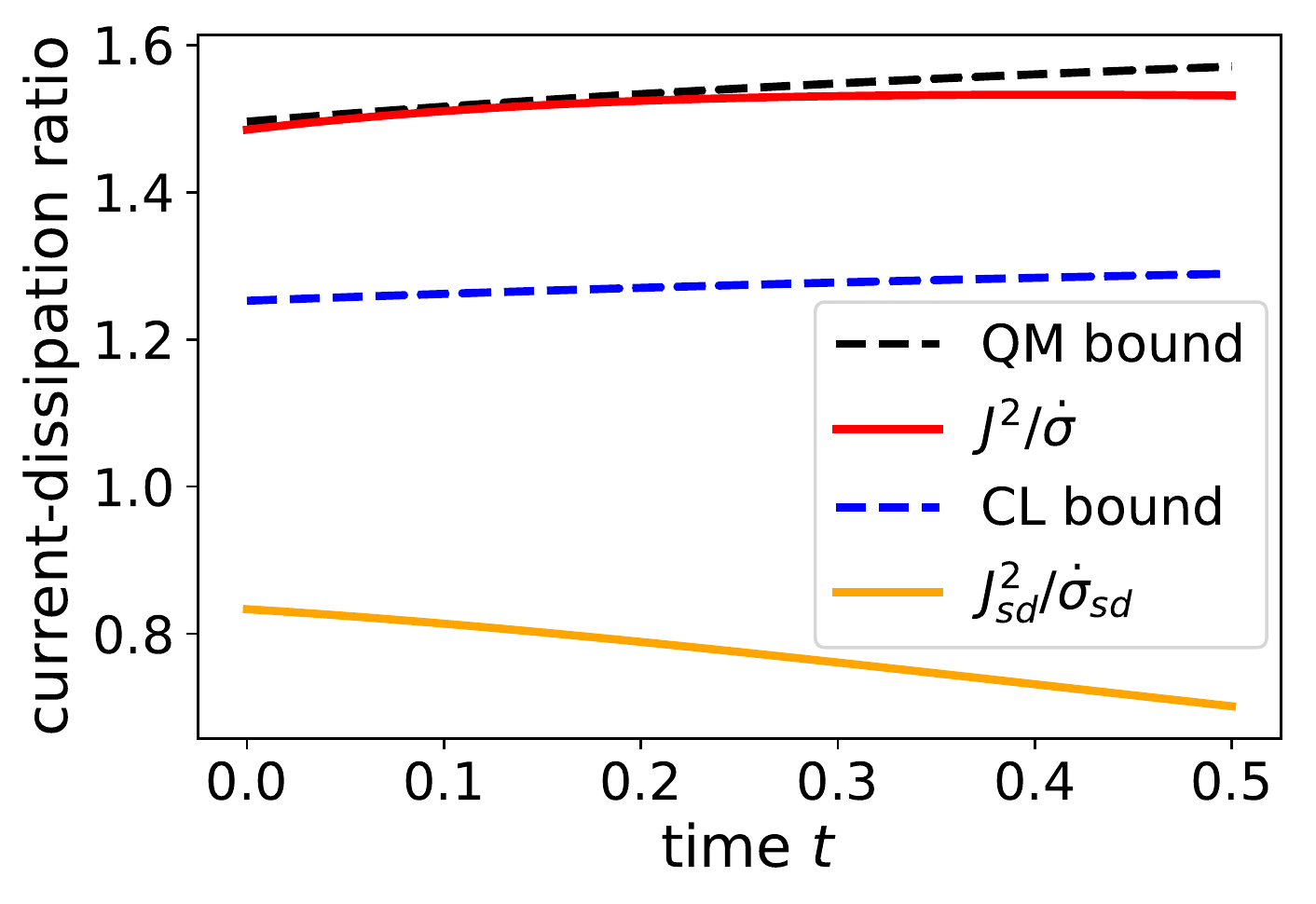}
\caption{Numerical check of the current-dissipation trade-off inequalities~\eqref{gosd} and~\eqref{go} during the heat engine cycle, step 1. Red and orange solid curves are the current-dissipation ratio $J^{2}/\dot{\sigma}$ for the states $\rho$ and $\rho_{\rm sd}$, respectively. Black dashed curve is the quantum bound $(A_{\rm cl}+A_{\rm qm})/2$ and the blue dashed curve is the classical bound $A_{\rm cl}/2$.  The parameters are $\omega_{\rm H}=2, \omega_{\rm C}=1, \beta_{\rm H}=0.6, \beta_{\rm C}=1.5, \tau_{\rm H}=0.5$ (Fig.~\ref{fig2} and Fig.~\ref{fig3}) and $\tau_{\rm C}=1.0$ (Fig.~\ref{fig2}).
}
\label{fig2}
\end{center}
\begin{center}
\includegraphics[width=.4\textwidth]{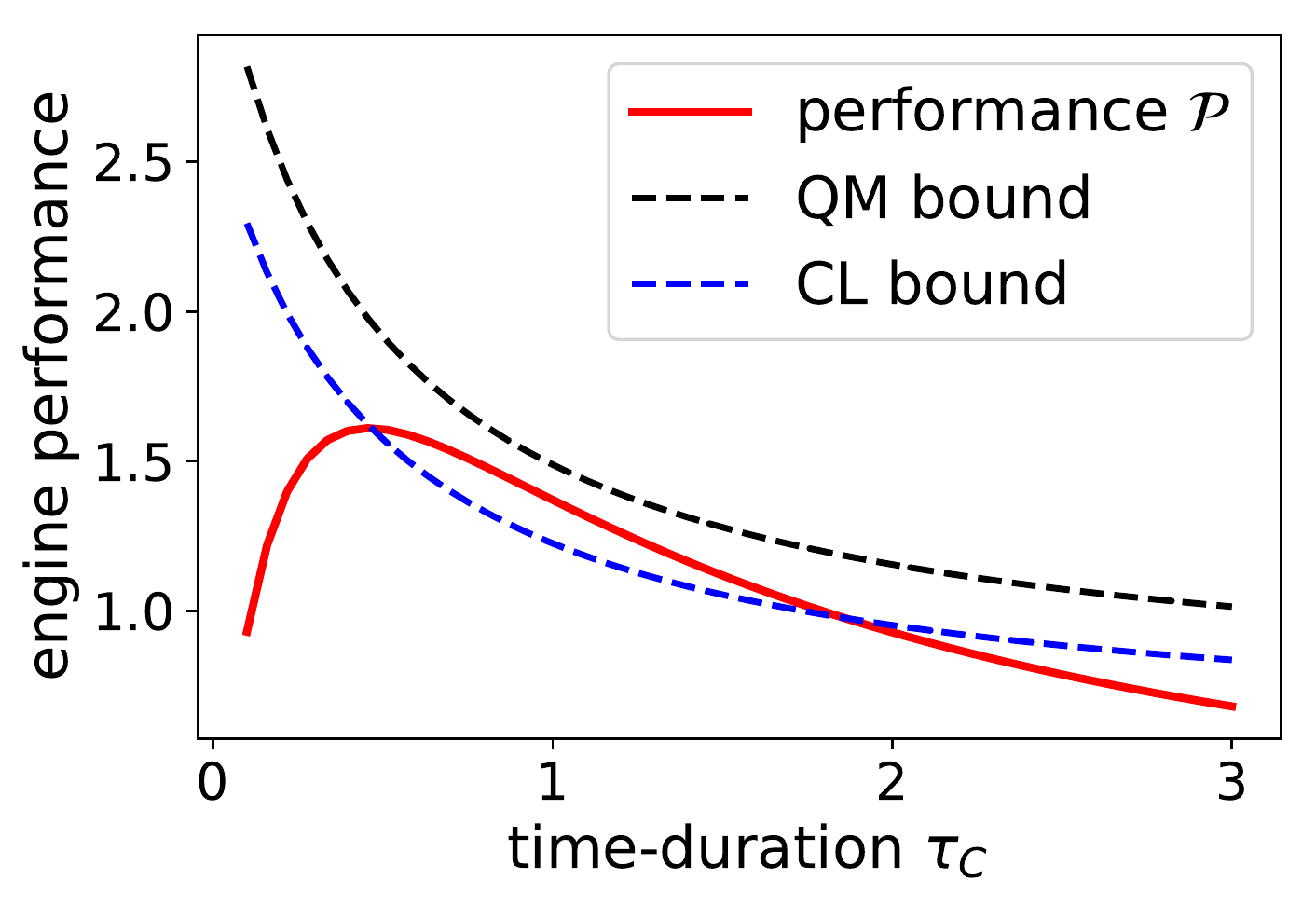}
\caption{Numerical calculation of the heat engine performance $\calP$ by varying the time-duration $\tau_{\rm C}$ of the heat engine cycle, step 3. Black dashed curve is the quantum bound $\bar{A}_{\rm cl}+\bar{A}_{\rm qm}$ and the blue dashed curve is the classical bound $\bar{A}_{\rm cl}$. Red solid curve shows the engine performance $\calP$, which exceeds the classical bound for some parameter range.
}
\label{fig3}
\end{center}
\end{figure}

As we have seen above, when the number of degeneracy is large, our $2N$-state model approximately achieves the Carnot efficiency with finite power. 
What if the number of degeneracy is small? Even in this case, our main results indicate interesting properties in the study of quantum heat engines. Here, instead of using the $2N$-state Hamiltonian~\eqref{2NHamiltonian} with $N=2$, we consider a two-qubit state superradiant model, since this model has been experimentally realized with superconducting qubits~\cite{SRrev,SRexp} (note that the qualitative behavior of the results does not change significantly between these two models).
We consider the heat engine cycle described above, and demonstrate the quantum advantage with numerical calculations. 
After the system relaxes to the stationary cycle, we calculate the heat current etc., and numerically check the inequalities~\eqref{gosd} and~\eqref{go} during step 1, plotted in Fig.~\ref{fig2}.
Clearly, the ratio $J(\rho)^2/\dot{\sigma}(\rho)$ exceeds the classical limit $A_{\cl}/2$ (note that $\rho=\rho_{\rm bd}$ holds in our example).
Finally, let us consider the power-efficiency trade-off relation by considering the following indicator of the engine \textit{performance}: $\calP:=(W/\tau)\times (\eta_{\rm Car}-\eta)^{-1}\times 2(2-\eta)^{2} \beta_{\rm C}^{-1}\eta^{-1}$.
The engine performance $\calP$ takes a large value when either the output power becomes large or the efficiency becomes close to the Carnot efficiency.
From \eqref{no-go}-\eqref{go}, we obtain two upper bounds on $\calP$ (
see Supplementary Information~\ref{p-e-t} for details): 
\begin{align}
\calP_{\cl}\le\bar{A}_{\cl}\enskip \mbox{and} \enskip\calP \le\bar{A}_{\cl}+\bar{A}_{\qm},\label{boundP}
\end{align}
where $\bar{A}_{\cl}$ and $\bar{A}_{\qm}$ are the time average of $A_{\cl}$ and $A_{\qm}$ per 1 engine cycle, and $\calP_{\cl}$ is the engine performance for $\rho_{\sd}$. Similar to \eqref{go}, when there is no coherence, the power and efficiency of the heat engines are bounded by $\bar{A}_{\cl}$. In this sense, $\bar{A}_{\cl}$ is the classical limitation on the performance of heat engines.
Meanwhile, when there exists coherence, a quantum heat engine can exceed the classical limitation up to $\bar{A}_{\qm}$.
With the 2-qubit superradiant model, we can numerically check that the power-efficiency performance of a quantum heat engine actually exceeds the classical limitation $\bar{A}_{\cl}$ for some parameter range, as shown in Fig.~\ref{fig3}.

Before concluding this paper, we give several comments and perspectives regarding the applications of our theoretical framework to the field of finite-time thermodynamics, photosynthesis, and quantum information theory. 

\textit{Finite-time Carnot engine:}
As a direct application of our main results, we gave a heat engine model which approximately attains the Carnot efficiency with finite power. Note that our strategy differs from previous studies, e.g., utilizing nonlinearity~\cite{Ponmurugan,Indekeu}, time-reversal symmetry breaking~\cite{Benenti,Mintchev}, specific system-bath coupling~\cite{Allahverdyan}, criticality with divergent energy fluctuations~\cite{Polettini,Campisi,Johnson}, and a large cycle time compared with the relaxation time~\cite{Ryabov}. In fact, our $2N$-state model does not use special properties of the dynamics, the cycle time is shorter than the relaxation time-scale, and the energy fluctuation remains $O(1)$. Note again that our strategy was to reduce dissipation via coherence. Therefore, we believe that our ``dissipation-less'' current-driven quantum heat engine adds new insight into the study of finite-time Carnot engine.

\textit{Quantitative understanding of the role of coherence in photosynthesis:}
An important issue in biology is the role of coherence in photosynthesis~\cite{Cheng-review,Engel,Mohseni,Ishizaki,Rebentrost,p-syn}, 
as recent experiments show that coherence actually survives for a sufficiently long time during the photosynthetic reaction~\cite{Cheng-review,Engel}.
Although there are many results reporting the effect of coherence in photosynthetic processes~\cite{Mohseni,Ishizaki,Rebentrost,p-syn}, there is no unified understandings about how the coherence actually contributes to a high light-harvesting efficiency performance.
Since theoretical models are often described by the quantum master equation \cite{Mohseni,Rebentrost}, it would be interesting to apply our framework to this problem and quantitatively clarify how coherence improves the energy transmission in photosynthesis compared with the classical bound. It would also be a interesting future direction to utilize our results and analyze the performance of biologically-inspired heat engine model of photocells~\cite{Chin,Scully}.

\textit{Difference between speakable and unspeakable coherence:}
Our results add new insight into the classification of coherence in quantum information theory, where
coherence is classified into two classes:
speakable coherence and unspeakable coherence~\cite{Marvian}.
Roughly speaking, speakable coherence refers to the coherence between bases that can be relabeled (e.g., computational basis in a quantum computer).
Conversely, unspeakable coherence refers to the coherence between bases that cannot be relabeled (e.g., energy eigenstates with different energy eigenvalues).
In our case, the difference between $\rho_{\bd}$ and $\rho_{\sd}$ reflects the difference between the speakable and unspeakable coherence in $\rho$, since the map $\rho\rightarrow\rho_{\bd}$ ($\rho\rightarrow\rho_{\sd}$) is the resource destroying map \cite{Liu} in the resource theory of asymmetry~\cite{Marvian-thesis} (coherence~\cite{Aberg,Winter}) which kills the unspeakable (speakable) coherence.
Therefore, our results give the following thermodynamic meanings to the classification of coherence: unspeakable coherence does not improve the performance of heat engines, and only the non-unspeakable part of the speakable coherence, quantified by $\calC_{l_1}(\rho_{\bd})$, contributes to the performance enhancement.

In this article, we gave a unified understanding of how quantum coherence affects the current-dissipation ratio.
Our results can be summarized in three basic rules as follows:
1. Coherence between different energy eigenspaces always reduces the ratio.
2. Coherence among degenerate states can be used to increases the ratio.
3. If there is enough coherence among degeneracy, the heat current can become macroscopic order while dissipation remains at constant order, realizing a  ``dissipation-less'' current. 
From the above observations, we clarified which type of quantum coherence contributes to the performance of heat engines. We have demonstrated this quantum enhancement by the 2-qubit system example, where the current-dissipation ratio and the engine performance exceed the classical bound. 
In addition, by utilizing the dissipation-less current, we have constructed a heat engine model which effectively attains the Carnot efficiency with finite-power.  
It is noteworthy to point out that our dissipation-less current induced by coherence resembles the superconducting current without energy dissipation, induced by large off-diagonal components. Our method is applicable to the energy flow caused by a chemical potential difference, and it may give stochastic thermodynamics viewpoint of the current-dissipation relation in superconducting phenomena.
We expect that our findings will further contribute to the understandings and design of low-dissipative energy transporting mechanisms in energy science, biology, and condensed matter physics.   
$\\$

\textbf{Acknowledgements:} 
We thank H. Hayakawa and Y. Hino for discussions and helpful comments.
We acknowledge support from the JSPS KAKENHI (H.T. and K.F.; Grant Number JP19K14610(HT) and  JP18J00454(KF)), and the Foundational Questions Institute Fund, a donor advised fund of Silicon Valley Community Foundation (K.F.; Grant Number FQXi-IAF19-06). 
$\\$

\textbf{Author Contributions:} 
The main ideas and formulations were developed by both authors. 
H.T. proved the main technical claims, supported by K.F., and K.F. gave the numerical calculations. 
Both authors wrote the manuscript.
$\\$

\textbf{Competing Interests:}
The authors declare no competing financial interests.

\clearpage


\begin{widetext}
\begin{center}
{\large \bf Supplemental Material for \protect \\ 
``Superconducting-like heat current:\\
Effective cancellation of current-dissipation trade off by quantum coherence''}\\
\vspace*{0.3cm}
Hiroyasu Tajima$^{1}$ and Ken Funo$^{2}$ \\
\vspace*{0.1cm}
$^{1}${\small \em Graduate School of Informatics and Engineering,
The University of Electro-Communications,
1-5-1 Chofugaoka, Chofu, Tokyo 182-8585, Japan}
\\
$^{2}${\small \em Theoretical Physics Laboratory, RIKEN Cluster for Pioneering Reserach, Wako-shi, Saitama 351-0198, Japan}
\end{center}

\setcounter{equation}{0}
\setcounter{lemma}{1}
\setcounter{page}{1}
\renewcommand{\theequation}{S.\arabic{equation}}

The supplementary information is organized as follows.
In Sec.~\ref{go-and-no-go}, we discuss the coherence effect on the current-dissipation trade-off relation, and give explicit proofs of our main results \eqref{no-go}, \eqref{gosd}, and \eqref{go} in the main text.
In Sec.~\ref{N+1}, we present the details of our first example which realizes a ``dissipation-less'' current.
As explained in the main text, here we use $2N$-state model, and show how coherence can be used to realize an $O(N)$ heat current with an $O(1)$ entropy production rate. In particular, we show that a steady current with a constant-order entropy production rate can occur using this model. In addition, we show that this model is able to implement a heat engine cycle which approximately attains the Carnot efficiency with finite power. 
In Sec.~\ref{SM-numeric}, we give details of the two-qubit superradiant model that we use to numerically check our results. 
Finally, in Sec.~\ref{p-e-t}, we discuss coherence effect on the power-efficiency trade-off relation of heat engines.

\section{Coherence effect on the current-dissipation trade-off}\label{go-and-no-go}

To clarify how the quantum coherence affects the current-dissipation trade-off, we have given a no-go and a go theorems in the main text. 
For the convenience to the readers, we present these theorems again.
The no-go theorem is stated as follows:
\begin{stheorem}\label{st1}
For arbitrary $\rho$, the following inequality holds:
\begin{align}
\frac{J(\rho)^2}{\dot{\sigma}(\rho)}\le\frac{J(\rho_{\bd})^2}{\dot{\sigma}(\rho_{\bd})}\label{s-nogo}
\end{align}
\end{stheorem}
Therefore, when the system Hamiltonian has no degeneracy, the limitation of $J(\rho)^2/\dot{\sigma}(\rho)$ is not enhanced by quantum coherence.

On the other hand, when there is degeneracy, quantum coherence can give positive effects on the current-dissipation trade-off. We can see this through the following go theorem:
\begin{stheorem}\label{st2}
For arbitrary $\rho$, the following two inequalities hold:
\begin{align}
&\frac{J(\rho_{\sd})^2}{\dot{\sigma}(\rho_{\sd})}\le\frac{A_{\cl}}{2} , \label{st2sd} \\
&\frac{J(\rho_{\bd})^2}{\dot{\sigma}(\rho_{\bd})}\le\frac{A_{\cl}+A_{\qm}}{2} , \label{st2bd}
\end{align}
where $A_{\cl}$ and $A_{\qm}$ are non-negative quanitites defined as follows:
\begin{align}
A_{\cl}&:=\Tr[\rho_{\sd}X]\mbox{  and  }A_{\qm}:=\calC_{X}\calC_{l_1}(\rho_{\bd})\\
X&:=\sum_{a,\omega}\gamma_{a}(\omega)\omega^2L^{\dagger}_{a,\omega}L_{a,\omega}\\
\calC_{X}&:=\max_{e, j, j': j\ne j'}|\bra{e,j}X\ket{e,j'}|
\end{align}
Here $C_{l_1}(...)$ is the $l_1$-norm of coherence with respect to the eigenbasis of the Hamiltonian $\{\ket{e,j}\}$: $C_{l_1}(...):=\sum_{(e,j)\ne (e',j')}|\bra{e,j}...\ket{e',j'}|.$
\end{stheorem}

Before showing theorems~\ref{st1} and~\ref{st2}, we present the quantum master equation~\eqref{master} again for convenience:
\begin{align}
\frac{\partial \rho}{\partial t}&=\calL[\rho]:=-\frac{i}{\hbar}[H,\rho]+{\cal D}[\rho], \label{supLE} \\
{\cal D}[\rho]&:=\sum_{a,\omega}\gamma_{a}(\omega)\left[L_{a,\omega}\rho L^{\dagger}_{a,\omega}-\frac{1}{2}\{L^{\dagger}_{a,\omega}L_{a,\omega},\rho\}\right].
\end{align}
Readers should note that we have slightly changed the notation from the main text and consider multiple Lindblad jump operators labeled by $a$. For example, $a$ can take two values, where $a={\rm H}$ ($a={\rm C}$) describes the dissipative effect arising from the hot (cold) bath. The Lindblad jump operator $L_{a,\omega}$ satisfies the following properties: $[L_{a,\omega},H]=\hbar\omega L_{a,\omega}$, $[L^{\dagger}_{a,\omega}L_{a,\omega},H]=0$ and $L_{a,\omega}=L^{\dagger}_{a,-\omega}$. For simplicity, we mainly consider the case where the system is attached to a single heat bath whose inverse temperature is $\beta$. Then, the detailed balance condition is expressed as $\gamma_{a}(\omega)/\gamma_{a}(-\omega)=\exp(\beta\hbar\omega)$. 

We also present the definition of the heat current and the entropy production rate:
\begin{align}
J(\rho)&=\Tr[H\partial_{t}\rho]=\Tr[H\calD[\rho]], \label{sup_defJ} \\
\dot{\sigma}(\rho)&=-\Tr[\partial_{t}\rho\log\rho]-\beta J(\rho)= -\Tr[\mathcal{L}[\rho]\log\rho]-\beta J(\rho). \label{sup_defEP}
\end{align}
Note that $J$ and $\dot{\sigma}$ for states $\rho_{\rm bd}$ and $\rho_{\rm sd}$ are defined by the second equality in~\eqref{sup_defJ} and \eqref{sup_defEP}.

\begin{proofof}{Supplementary Theorem \ref{st1}}
We show this result by showing the following two relations:
\begin{align}
J(\rho)=J(\rho_{\mathrm{bd}})\label{J=}\\
\dot{\sigma}(\rho_{\mathrm{bd}})\le \dot{\sigma}(\rho)\label{sig<}
\end{align}
Clearly, if these two relations hold, then the inequality \eqref{s-nogo} also holds.
We first show \eqref{J=} as follows:
\begin{align}
J(\rho)&=\Tr[H\calD[\rho]]
=\Tr\left[H\sum_{a,\omega}\gamma_{a}(\omega)\left[L_{a,\omega}\rho L^{\dagger}_{a,\omega}-\frac{1}{2}\{L^{\dagger}_{a,\omega}L_{a,\omega},\rho\}\right]\right]\nonumber\\
&=-\sum_{a,\omega}\omega\gamma_{a}(\omega)\Tr[L^{\dagger}_{a,\omega}L_{a,\omega}\rho]\nonumber\\
&=-\sum_{a,\omega}\omega\gamma_{a}(\omega)\Tr[L^{\dagger}_{a,\omega}L_{a,\omega}\rho_{\bd}]=J(\rho_{\bd}).\label{J=proof}
\end{align}
Here we use $[L_{a,\omega},H]=\omega L_{a,\omega}$ in the second line and $[L^{\dagger}_{a,\omega}L_{a,\omega},\Pi_e]=0$, which is given by $[L^{\dagger}_{a,\omega}L_{a,\omega},H]=0$, in the third line.

Next, we show \eqref{sig<}.
By focusing on $\dot{\sigma}(\rho):=\dot{S}(\rho)-\beta J(\rho)$ and $J(\rho)=J(\rho_{\bd})$, the inequality \eqref{sig<} is equivalent to the inequality $\dot{S}(\rho)-\dot{S}(\rho_{\bd})\ge0$.
To derive this inequality, we firstly convert $\dot{S}(\rho)-\dot{S}(\rho_{\bd})$ as follows:
\begin{align}
\dot{S}(\rho)-\dot{S}(\rho_{\bd})&=\lim_{dt\rightarrow0}\frac{1}{dt}\Tr[-\rho(t+dt)\log\rho(t+dt)+\rho(t)\log\rho(t)+\rho_{\bd}(t+dt)\log\rho_{\bd}(t+dt)-\rho_{\bd}(t)\log\rho_{\bd}(t)]\nonumber\\
&=\lim_{dt\rightarrow0}\frac{1}{dt}\Tr[-\rho(t+dt)\log\rho(t+dt)+\rho(t)\log\rho(t)+\rho(t+dt)\log\rho_{\bd}(t+dt)-\rho(t)\log\rho_{\bd}(t)]\nonumber\\
&=\lim_{dt\rightarrow0}\frac{1}{dt}(D(\rho(t)\|\rho_{\bd}(t))-D(\rho(t+dt)\|\rho_{\bd}(t+dt))).\label{S12}
\end{align}
Here we define $\rho_{\rm bd}(t+dt)$ as
\begin{align}
\rho_{\rm bd}(t+dt)&:=\rho_{\bd}+dt\calL[\rho_{\bd}],
\end{align}
where the map $\calL$ is defined in Eq.~\eqref{supLE}.
Since $\rho_{\bd}$ commutes with $H$, $\rho_{\bd}(t+dt)$ can be expressed as follows:
\begin{align}
\rho_{\rm bd}(t+dt)&=\rho_{\rm bd}(t)+dt \mathcal{D}[\rho_{\rm bd}(t)] \nonumber \\
&=\sum_{n}\Pi_{n}(t) [\rho(t)+dt \mathcal{D}[\rho(t)]]\Pi_{n}(t) \nonumber \\
&=\sum_{n}\Pi_{n}(t)\rho(t+dt)\Pi_{n}(t),
\end{align}
where the second line is obtained because the Lindblad operators let the system jump from one eigenspace ($\Pi_{n}$) to another ($\Pi_{m}$). As a result, we have
\begin{align}
\Tr[\rho_{\rm bd}(t+dt)\ln\rho_{\rm bd}(t+dt)]=\Tr[\rho(t+dt)\ln\rho_{\rm bd}(t+dt)]
\end{align}
and 
\begin{align}
D(\rho(t+dt)||\rho_{\rm bd}(t+dt))=D(\Lambda[\rho(t)]||\Lambda[\rho_{\rm bd}(t)]), \label{sup_dCPTP}
\end{align}
where
\begin{align}
\Lambda[\bullet]=(1+dt\mathcal{L})\bullet
\end{align}
is a CPTP map which describes an infinitesimal time-evolution generated by the Lindblad master equation. By using the monotonicity of the relative entropy~\cite{Nielsen}, we have
\begin{align}
D(\rho(t)||\rho_{\rm bd}(t))\geq D(\Lambda[\rho(t)]||\Lambda[\rho_{\rm bd}(t)]). \label{sup_mono}
\end{align}
We finally combine \eqref{S12}, \eqref{sup_dCPTP} and \eqref{sup_mono} and prove $\dot{S}(\rho)-\dot{S}(\rho_{\rm bd})\geq 0$. 
\end{proofof}

\begin{proofof}{Supplementary Theorem \ref{st2}}
We first decompose $\rho_{\bd}$ as $\rho_{\bd}=\sum_np_n\ket{n}\bra{n}$.  Note that $\rho_{\bd}$ is block-diagonalized in the energy eigenspace, each $\ket{n}$ is an eigenstate of $H$.
For this basis $\{\ket{n}\}$, we define $W^{\omega,a}_{m,n}:=\gamma_{a}(\omega)|\bra{m}L_{\omega,a}\ket{n}|^2$, which can be interpreted as a transition rate from $\ket{n}$ to $\ket{m}$ induced by the bath.
By following Ref.~\cite{Funo-QSL} and using the third line of \eqref{J=proof}, we have
\begin{align}
J(\rho_{\bd})&=-\sum_{a,\omega,m,n}\omega W^{\omega,a}_{m,n}p_n\nonumber\\
&=-\frac{1}{2}\sum'_{a,\omega,m,n}\omega(W^{\omega,a}_{m,n}p_n-W^{-\omega,a}_{n,m}p_m),\label{15}
\end{align}
where $\sum'$ is the summation excluding $(\omega=0)\land(m=n)$. Similarly, the entropy production rate can be expressed as~\cite{Funo-QSL}
\begin{align}
\dot{\sigma}(\rho_{\bd})&=-\sum_{a,\omega}\gamma_{a}(\omega)\Tr[L_{a,\omega}\rho L^{\dagger}_{a,\omega}\ln\rho_{\bd}]+\sum_{a,\omega}\gamma_{a}(\omega)\Tr[L^{\dagger}_{a,\omega}L_{a,\omega}\rho_{\bd}\ln\rho_{\bd}] -\beta J(\rho_{\rm bd}) \nonumber \\
&=-\sum_{a,\omega,m,n} W^{\omega,a}_{m,n}p_n\ln p_{m} + \sum_{a,\omega,m,n} W^{\omega,a}_{m,n}p_n\ln p_{n} + \sum_{a,\omega,m,n}\beta\omega W^{\omega,a}_{m,n}p_n\nonumber\\
&=\sum'_{a,\omega,m,n} W^{\omega,a}_{m,n}p_n \log \frac{ W^{\omega,a}_{m,n}p_n}{W^{-\omega,a}_{n,m}p_m}\geq 0,\label{15EP}
\end{align}
by noting that $W^{\omega,a}_{m,n}=e^{\beta\omega}W^{-\omega,a}_{n,m}$. The last inequality results from the nonegativity of the relative entropy~\cite{Nielsen}. By using above, we evaluate the absolute value of the heat current as follows:
\begin{align}
|J(\rho_{\bd})|&=\Biggl|\sum'_{a,\omega,m,n}\frac{\omega}{2}(W^{\omega,a}_{m,n}p_n-W^{-\omega,a}_{n,m}p_m)\Biggr|=\Biggl|\sum'_{a,\omega,m,n}\frac{\omega}{2}\sqrt{W^{\omega,a}_{m,n}p_n+W^{-\omega,a}_{n,m}p_m}\frac{(W^{\omega,a}_{m,n}p_n-W^{-\omega,a}_{n,m}p_m)}{\sqrt{W^{\omega,a}_{m,n}p_n+W^{-\omega,a}_{n,m}p_m}} \Biggr| \nonumber\\
&\le\sqrt{\sum'_{a,\omega,m,n}\frac{\omega^2}{4}(W^{\omega,a}_{m,n}p_n+W^{-\omega,a}_{n,m}p_m)}\sqrt{\sum'_{a,\omega,m,n}\frac{(W^{\omega,a}_{m,n}p_n-W^{-\omega,a}_{n,m}p_m)^2}{W^{\omega,a}_{m,n}p_n+W^{-\omega,a}_{n,m}p_m}}\nonumber\\
&\le\sqrt{\sum'_{a,\omega,m,n}\frac{\omega^2}{4}(W^{\omega,a}_{m,n}p_n+W^{-\omega,a}_{n,m}p_m)}\sqrt{\frac{1}{2}\sum'_{a,\omega,m,n}(W^{\omega,a}_{m,n}p_n-W^{-\omega,a}_{n,m}p_m)\log\frac{W^{\omega,a}_{m,n}p_n}{W^{-\omega,a}_{n,m}p_m}}\nonumber\\
&=\sqrt{\sum'_{a,\omega,m,n}\frac{\omega^2}{4}(W^{\omega,a}_{m,n}p_n+W^{-\omega,a}_{n,m}p_m)}\sqrt{\sum'_{a,\omega,m,n}W^{\omega,a}_{m,n}p_n\log\frac{W^{\omega,a}_{m,n}p_n}{W^{-\omega,a}_{n,m}p_m}}\nonumber\\
&=\sqrt{\sum'_{a,\omega,m,n}\frac{\omega^2}{2}W^{\omega,a}_{m,n}p_n}\sqrt{\dot{\sigma}(\rho_{\bd})}=\sqrt{\sum_{a,\omega,m,n}\frac{\omega^2}{2}W^{\omega,a}_{m,n}p_n}\sqrt{\dot{\sigma}(\rho_{\bd})} \label{JBbound1}
\end{align}
Here we use the Cauchy-Schwartz inequality in the second line and the inequality $\frac{(a-b)^2}{a+b}\le \frac{a-b}{2}\log\frac{a}{b}$ in the third line.
We further use the following inequality
\begin{align}
\sum_{a,\omega,m,n}\frac{\omega^2}{2}W^{\omega,a}_{m,n}p_n
&=
\Tr[\sum_{\omega,a}\frac{\omega^2}{2}\gamma_{a}(\omega)L_{\omega,a}\rho_{\bd}L^{\dagger}_{\omega,a}]\nonumber\\
&=\frac{1}{2}\Tr[X\rho_{\bd}]\nonumber\\
&=\frac{\Tr[X\rho_{\sd}]+\Tr[X(\rho_{\bd}-\rho_{\sd})]}{2}\nonumber\\
&\le \frac{\Tr[X\rho_{\sd}]+\calC_{X}\calC_{l_1}(\rho_{\bd})}{2}\nonumber\\
&=\frac{A_{\cl}+A_{\qm}}{2}, \label{JBbound2}
\end{align}
where in the forth line, we use H\"{o}lder's inequality 
\beq
\sum_{k} |x_{k}y_{k}| \leq \Bigl(\sum_{k}|x_{k}|^{p} \Bigr)^{1/p}\Bigl(\sum_{k}|y_{k}|^{q} \Bigr)^{1/q},
\eeq
with $p=\infty$, $q=1$, $x_{k}=\langle e,j|X|e',j'\rangle$, $y_{k}=\langle e',j'|\rho|e,j\rangle$, and $k\in\{e,j,e',j'|(e,j)\neq (e',j')\}$. We also use the relation $\max_{k}|x_{k}|=\max_{e, j, j': j\ne j'}|\bra{e,j}X\ket{e,j'}|=\calC_{X}$ since $\langle e,j|X|e',j'\rangle=0$ for $e\neq e'$. By combining \eqref{JBbound1} and \eqref{JBbound2}, we obtain the desired result~\eqref{st2bd}. The derivation of~\eqref{st2sd} can be done in a similar manner.  
The non-negativity of $A_{\cl}$ is obvious from the non-negativity of the operator $X$.
\end{proofof}

\section{heat current without dissipation: $2N$-state model}\label{N+1}

In this section, we show the detailed analysis on the $2N$-state model.
In this model, the system Hamiltonian  and the interaction Hamiltonian between the system and the heat bath are given by
\begin{align}
H=\sum^{N}_{j=1}\hbar\omega\ket{{\rm e},j}\bra{{\rm e},j},\enskip\mbox{ and }\enskip
H_{\mathrm{int}}=\sum^{N}_{j,j'=1}(\sigma^{j,j'}_{+}\otimes B
+\sigma^{j,j'}_{-}\otimes B^{\dagger}).
\end{align}
Here $\ket{{\rm g},j}$ is the $j$-th degenerated ground state, and the state $\ket{{\rm e},j}$ is the $j$-th degenerated excited state, meaning that the ground state energy and the excited state energy are both $N$-degenerate, and the total number of states is given by $2N$.  
Also, $\sigma^{j,j'}_{+}:=\ket{{\rm e},j}\bra{{\rm g},j'}$, $\sigma^{j,j'}_{-}:=\ket{{\rm g},j}\bra{{\rm e},j'}$ and $B$ is a Hermitian operator of the bath.
After taking the standard weak-coupling, Born-Markov, and rotating-wave approximations, the time-evolution of the system is described by the quantum master equation \eqref{master}, where we denote $\Gamma_{\downarrow}=\gamma(\omega)$, $\Gamma_{\uparrow}=\gamma(-\omega)$, and $L=L_{\omega}=L^{\dagger}_{-\omega}$ to simplify the notations.  
Then, the master equation is written as
\begin{align}
\frac{\partial \rho}{\partial t}&=-\frac{i}{\hbar}[H,\rho]+{\cal D}[\rho],\label{s-master}\\
{\cal D}[\rho]&=\Gamma_{\downarrow}\left[L\rho L^{\dagger}-\frac{1}{2}\{L^{\dagger}L,\rho\}\right]+\Gamma_{\uparrow}\left[L^{\dagger}\rho L-\frac{1}{2}\{LL^{\dagger},\rho\}\right],
\end{align}
where 
\beq
L:=\sum_{j,j'} \sigma^{j,j'}_{-} \label{defL}
\eeq
is the Lindblad operator that describes a correlated decay. Here, the detailed balance relation between the transition rates is expressed as $\Gamma_{\downarrow}/\Gamma_{\uparrow}=\exp[\beta\hbar\omega]$.

\subsection{Instanteneous heat current in $2N$-state model}

In this subsection, we give an example which gives an $O(N)$ heat current with an $O(1)$ entropy production rate. 
As we explained in the main text, we take the state $\rho^{+}:=p_{\rm g}\ket{{\rm g},+}\bra{{\rm g},+}+p_{\rm e}\ket{{\rm e},+}\bra{{\rm e},+}.$
In what follows, we show that by properly setting the probability $p_{\rm g}$ and $p_{\rm e}$, we can obtain the dissipation-less current.

We firstly note that $\calD[\ket{{\rm g},+}\bra{{\rm g},+}]\propto\ket{{\rm e},+}\bra{{\rm e},+}$ and $\calD[\ket{{\rm e},+}\bra{{\rm e},+}]\propto\ket{{\rm g},+}\bra{{\rm g},+}$, and thus when we start with the initial state $\rho(0)=\rho^{+}$, the state $\rho(t)$ at time $t$ always takes the following form: $\rho(t)=p_{\rm g}(t)\ket{{\rm g},+}\bra{{\rm g},+}+p_{\rm e}(t)\ket{{\rm e},+}\bra{{\rm e},+}$.
Therefore, the time-evolution of the system is fully determined by the following set of equations:
\begin{align}
\partial_{t}p_{\rm g}(t)&=N^2\Gamma_{\downarrow}p_{\rm e}(t)-N^2\Gamma_{\uparrow}p_{\rm g}(t),\\
\partial_{t}p_{\rm e}(t)&=-N^2\Gamma_{\downarrow}p_{\rm e}(t)+N^2\Gamma_{\uparrow}p_{\rm g}(t),
\end{align}
and the heat current and the entropy production rate take the following forms:
\begin{align}
J(\rho(t))&=N^2\hbar\omega(\Gamma_{\uparrow}p_{\rm g}(t)-\Gamma_{\downarrow}p_{\rm e}(t)),\\
\dot{\sigma}(\rho(t))&=N^2(\Gamma_{\uparrow}p_{\rm g}(t)-\Gamma_{\downarrow}p_{\rm e}(t))\log\frac{\Gamma_{\uparrow}p_{\rm g}(t)}{\Gamma_{\downarrow}p_{\rm e}(t)} .
\end{align}

Now, we set $p_{\rm g}$ and $p_{\rm e}$ such that they satisfy the relation
\begin{align}
p_{\rm g}=(1+a_N)p_{\rm e}e^{\beta\hbar\omega} \Leftrightarrow \Gamma_{\uparrow}p_{0}=(1+a_N)\Gamma_{\downarrow}p_{1}, \label{ONpge}
\end{align}
where we have introduced a number $a_N$ of the order of $O(1/N)$. 
The above choice~(\ref{ONpge}) indeed allows us to obtain the dissipation-less current: 
\begin{align}
J(\rho^{+})&=N^2a_N\Gamma_{\downarrow}p_{\rm e}\hbar\omega\nonumber\\
&=O(N),\\
\dot{\sigma}(\rho^{+})&=N^2a_N\log(1+a_N) \Gamma_{\downarrow}p_{\rm e} \approx N^2a^2_N \Gamma_{\downarrow}p_{\rm e} \nonumber\\
&=O(1).
\end{align}

\subsection{$A_{\qm}$ and $A_{\cl}$ in the $2N$-state model}\label{2N-A}

The scaling behavior of $A_{\qm}$ and $A_{\cl}$ can be quantitatively different in the $2N$-state model, since $A_{\qm}$ can scale up to $O(N^{2})$, but $A_{\cl}$ can only scale up to $O(N)$.  
To see this scaling behavior, we first note that the operator $X$ takes the form
\begin{align}
X=\hbar^2\omega^2\Gamma_{\uparrow} N\sum_{j,j'}\ket{{\rm g},j}\bra{{\rm g},j'}+\hbar^2\omega^2 \Gamma_{\downarrow} N\sum_{j,j'}\ket{{\rm e},j}\bra{{\rm e},j'}.
\end{align}
We also note that for an arbitrary state $\rho$, the decohered state $\rho_{\sd}$ is written as $\rho_{\sd}=\sum_{j}p_{{\rm g},j}\ket{{\rm g},j}\bra{{\rm g},j}+\sum_{j'}p_{{\rm e},j'}\ket{{\rm e},j'}\bra{{\rm e},j'}$, where $p_{k,j}$ is some probability distribution.
Therefore, 
\begin{align}
A_{\cl}=\Tr[X\rho_{\sd}]=\hbar^2\omega^2 N \Gamma_{\downarrow}\sum_{j}p_{{\rm g},j}+\hbar^2\omega^2 N \Gamma_{\uparrow}\sum_{j}p_{{\rm e},j} \leq \hbar^2\omega^2 N (\Gamma_{\downarrow}+\Gamma_{\uparrow})=O(N),
\end{align}
indicating that $A_{\rm cl}$ can scale at most linearly in terms of $N$. On the other hand, for a state $\rho^{+}:=p_{{\rm g}}\ket{{\rm g},+}\bra{{\rm g},+}+p_{{\rm e}}\ket{{\rm e},+}\bra{{\rm e},+}$ with $p_{{\rm g}}$ and $p_{{\rm e}}$ being arbitrarily, $A_{\rm qm}$ scales quadratically in terms of $N$:
\begin{align}
A_{\qm}=\calC_{X}\calC_{l_1}(\rho^+)=\hbar^2\omega^2\Gamma_{\downarrow} N\times \frac{N^2-N}{N}=O(N^2).
\end{align}

\subsection{Carnot efficiency with finite power in the $2N$-state model}\label{Carnot2N}

In this subsection, we give details of our cyclic heat engine which approximately attains the Carnot efficiency with finite power.
We prepare two heat baths whose inverse temperatures are $\beta_H$ and $\beta_C$, respectively.
We denote the energy gaps $\hbar\omega^{(N)}_H$ and $\hbar\omega_C$ when the system is connected to the hot bath and the cold bath, respectively. We require those energy gaps to satisfy the following relation:
\begin{align}
\hbar(\beta_C\omega_C-\beta_H\omega^{(N)}_H)=\log(1+a_N)\approx a_N=O(1/N)\label{18}, 
\end{align}
where $a_N=O(1/N)$ is a parameter which describes the speed of our control. Here, we first fix $\omega_{\rm C}$ and then define the $N$-dependent frequency $\omega^{(N)}_H$ via Eq.~\eqref{18} for convenience. From Eq.\eqref{18}, we have
\begin{align}
1+a_N=e^{\hbar(\beta_C\omega_C-\beta_H\omega^{(N)}_H)}\label{aCH}.
\end{align}

In what follows, we explain the details of our four-step cycle engine: 
\begin{description}
\item{Step 0}: At beginning, the system is connected to the cold heat bath. The system Hamiltonian and the dissipator are given as follows:
\begin{align}
H_C&=\hbar\omega_C\sum^{N}_{j=1}\ket{{\rm e},j}\bra{{\rm e},j}\\
{\cal D}_C[\rho]&=\Gamma^{C}_{\downarrow}\left[L\rho L^{\dagger}-\frac{1}{2}\{L^{\dagger}L,\rho\}\right]+\Gamma^{C}_{\uparrow}\left[L^{\dagger}\rho L-\frac{1}{2}\{LL^{\dagger},\rho\}\right]
\end{align}
where $L$ is given in Eq.~\eqref{defL}, and the transition rates are assumed to satisfy the detailed balance relation: $\Gamma^{C}_{\downarrow}/\Gamma^{C}_{\uparrow}=e^{\beta_C\hbar\omega_C}$. 
We set the initial state as
\begin{align}
\rho'^{C}_{ss}&=\rho'^{C}_{ss,{\rm gg}}\ket{{\rm g},+}\bra{{\rm g},+}+\rho'^{C}_{ss,{\rm ee}}\ket{{\rm e},+}\bra{{\rm e},+},\\
\Gamma^{C}_{\uparrow}\rho'^{C}_{ss,{\rm gg}}&=\frac{1}{1+0.45a_N}\Gamma^{C}_{\downarrow}\rho'^{C}_{ss,{\rm ee}}.
\end{align}
Note that this state is not the steady state. 
However, as we will see later, the total cycle becomes a steady cycle.
\item{Step 1}: We turn off the interaction between the system and the cold bath, and change the system Hamiltonian to the following one instantaneously:
\begin{align}
H^{(N)}_H&=\hbar\omega^{(N)}_H\sum^{N}_{j=1}\ket{{\rm e},j}\bra{{\rm e},j}
\end{align}
The state of the system is unchanged during this sudden quench process, since the Hamiltonian before and after the quench is commutative.
\item{Step 2}: We turn on the interaction between the system and the hot bath.
The dissipator is given as
\begin{align}
{\cal D}_H[\rho]&=\Gamma^{H,(N)}_{\downarrow}\left[L\rho L^{\dagger}-\frac{1}{2}\{L^{\dagger}L,\rho\}\right]+\Gamma^{H,(N)}_{\uparrow}\left[L^{\dagger}\rho L-\frac{1}{2}\{LL^{\dagger},\rho\}\right],
\end{align}
where $L$ is again given by Eq.~\eqref{defL}. 
Due to the detailed balance, $\Gamma^{H,(N)}_{\downarrow}/\Gamma^{H,(N)}_{\uparrow}=e^{\beta_H\hbar\omega^{(N)}_H}$ holds.
The initial state of the step 2 is $\rho'^{C}_{ss}$.
Due to \eqref{aCH}, the relation $\Gamma^{C}_{\downarrow}/\Gamma^{C}_{\uparrow}=(1+a_N)\Gamma^{H,(N)}_{\downarrow}/\Gamma^{H,(N)}_{\uparrow}$ holds, and thus the state $\rho'^{C}_{ss}$ satisifies
\begin{align}
\Gamma^{H,(N)}_{\uparrow}\rho'^{C}_{ss,{\rm gg}}=\frac{(1+a_N)}{1+0.45a_N}\Gamma^{H,(N)}_{\downarrow}\rho'^{C}_{ss,{\rm ee}}\label{25}.
\end{align}
Then, we wait until the state becomes the following $\rho'^{H}_{ss}$
\begin{align}
\rho'^{H}_{ss}&=\rho'^{H}_{ss,{\rm gg}}\ket{{\rm g},+}\bra{{\rm g},+}+\rho'^{H}_{ss,{\rm ee}}\ket{{\rm e},+}\bra{{\rm e},+},\\
\Gamma^{H,(N)}_{\uparrow}\rho'^{H}_{ss,{\rm gg}}&=(1+0.45a_N)\Gamma^{H,(N)}_{\downarrow}\rho'^{H}_{ss,{\rm ee}}\label{27}
\end{align}
Since we do \textit{not} wait until the system is completely thermalized, the process time of this step is finite.
Let us evaluate how long this step takes.
When a state diagonalized with $\ket{{\rm g},+}$ and $\ket{{\rm e},+}$ satisifies
\begin{align}
\Gamma^{H,(N)}_{\uparrow}\rho_{\rm gg}=(1+sa_N)\Gamma^{H,(N)}_{\downarrow}\rho_{\rm ee}
\end{align}
the state satisifies
\begin{align}
\partial_t \rho_{\rm gg}=-sN^2a_N\\
\partial_t \rho_{\rm ee}=+sN^2a_N
\end{align}
Consequently, the speed of the state transformation is equal to $0.55N^2a_N=O(N)$ at the beginning of the step 2 and equal to $0.45N^2a_N=O(N)$ at the end of the step 2. 
Therefore, the process time of the step 2 is $O(1/N^2)$.
Also, in this step, the entropy production rate is always $O(1)$.
Therefore, the entropy production in this step is $O(1/N^2)$.

It is noteworthy that the process time of this step is shorter than the half of the relaxation time. The reason is that the step 2 starts with $\rho'^{C}_{ss}$ satisfying $\Gamma^{H,(N)}_{\uparrow}\rho'^{C}_{ss,{\rm gg}}\approx(1+0.55a_N)\Gamma^{H,(N)}_{\downarrow}\rho'^{C}_{ss,{\rm ee}}$ and finishes with $\rho'^{H}_{ss}$ satisfying $\Gamma^{H,(N)}_{\uparrow}\rho'^{H}_{ss,{\rm gg}}=(1+0.45a_N)\Gamma^{H,(N)}_{\downarrow}\rho'^{H}_{ss,{\rm ee}}$.
When a state $\rho$ satisfies $\Gamma^{H,(N)}_{\uparrow}\rho_{\rm gg}\approx(1+0.55a_N)\Gamma^{H,(N)}_{\downarrow}\rho_{++}$, the state $\rho$ is the steady state.
Therefore, a state satisfying $\Gamma^{H,(N)}_{\uparrow}\rho_{\rm gg}=(1+sa_N)\Gamma^{H,(N)}_{\downarrow}\rho_{++}$ becomes another state $\rho'$ satisfying $\Gamma^{H,(N)}_{\uparrow}\rho'_{00}=(1+s'a_N)\Gamma^{H,(N)}_{\downarrow}\rho'_{++}$ where $s/s'=e^{-1}$.
Therefore, since $0.45/0.55>e^{-1/2}$, the process time of the step 2 is shorter than the half of the relaxation time.
\item{Step 3} We turn off the interaction between the system and the hot bath, and change the Hamiltonian to $H_C$. At the end of this step, the state of the system is $\rho'^{H}_{ss}$.
\item{Step 4} We connect the system to the cold bath. Then, due to \eqref{aCH}, the following relation holds
\begin{align}
\Gamma^{C}_{\uparrow}\rho'^{H}_{ss,{\rm gg}}=\frac{(1+0.45a_N)}{1+a_N}\Gamma^{C}_{\downarrow}\rho'^{H}_{ss,{\rm ee}}.
\end{align}
We wait until the state becomes $\rho'^{C}_{ss}$.
In the same way as the step 2, we obtain that the process time is $O(1/N^2)$, and the entropy production is $O(1/N^2)$, and the process time is shorter than the half of the relaxation time.
\end{description}
So far, we have seen the entropy production is $O(1/N^2)$, and the cycle time is $O(1/N^2)$.
Let us evaluate the average power of this cycle.
The work amount is
\begin{align}
W&=\Tr[(H_C-H^{(N)}_H)\rho'^{C}_{ss}]+\Tr[(H^{(N)}_H-H_C)\rho'^{H}_{ss}]\nonumber\\
&=(\hbar\omega^{(N)}_H-\hbar\omega_C)(\rho'^{H}_{ss,{\rm ee}}-\rho'^{C}_{ss,{\rm ee}})\label{workcarnot}
\end{align}
Due to \eqref{18},
\begin{align}
\hbar\omega^{(N)}_H-\hbar\omega_C=(1-\frac{\beta_H}{\beta_C})\hbar\omega^{(N)}_H-\frac{a_N}{\beta_C}=O(1)\label{keyefficiency}
\end{align}
Due to \eqref{25} and \eqref{27},
\begin{align}
\frac{\rho'^{C}_{ss,{\rm ee}}}{\rho'^{C}_{ss,{\rm gg}}}&=\frac{(1+0.45a_N)^2}{1+a_N}\frac{\rho'^{H}_{ss,{\rm ee}}}{\rho'^{H}_{ss,{\rm gg}}}\nonumber\\
&\approx(1-0.1a_N)\frac{\rho'^{H}_{ss,{\rm ee}}}{\rho'^{H}_{ss,{\rm gg}}}
\end{align}
With using the sum of the probability is 1,
\begin{align}
\frac{\rho'^{C}_{ss,{\rm ee}}}{1-\rho'^{C}_{ss,{\rm ee}}}&\approx(1-0.1a_N)\frac{\rho'^{H}_{ss,{\rm ee}}}{1-\rho'^{H}_{ss,{\rm ee}}}
\end{align}
Multiplying the both side by $(1-\rho'^{C}_{ss,{\rm ee}})(1-\rho'^{H}_{ss,{\rm ee}})$, we obtain
\begin{align}
\rho'^{C}_{ss,{\rm ee}}-\rho'^{H}_{ss,{\rm ee}}\rho'^{C}_{ss,{\rm ee}}\approx\rho'^{H}_{ss,{\rm ee}}-\rho'^{H}_{ss,{\rm ee}}\rho'^{C}_{ss,{\rm ee}}-0.1a_N\rho'^{H}_{ss,{\rm ee}}+0.1a_N\rho'^{H}_{ss,{\rm ee}}\rho'^{C}_{ss,{\rm ee}}
\end{align}
Therefore,
\begin{align}
\rho'^{H}_{ss,{\rm ee}}-\rho'^{C}_{ss,{\rm ee}}\approx 0.1a_N\rho'^{H}_{ss,{\rm ee}}(1-\rho'^{C}_{ss,{\rm ee}})=O(1/N).
\end{align}
Consequently, the order of the work amount is $O(1/N)$.
Since the cycle time is $O(1/N^2)$, the average power is $O(N)$.

Finally, let us check the efficiency of the cycle.
The heat amount from the hot bath is given as
\begin{align}
Q&=\Tr[H^{(N)}_H(\rho'^{H}_{ss}-\rho'^{C}_{ss})]\nonumber\\
&=\hbar\omega^{(N)}_H(\rho'^{H}_{ss,{\rm ee}}-\rho'^{C}_{ss,{\rm ee}})\label{heatcarnot}
\end{align}
Due to \eqref{workcarnot} and \eqref{keyefficiency},
\begin{align}
\eta=\frac{\omega^{(N)}_H-\omega_C}{\omega^{(N)}_H}=\eta_{\mathrm{Car}}-\frac{a_N}{\beta_C\hbar\omega^{(N)}_H}=\eta_{\mathrm{Car}}-O(1/N).
\end{align}
Therefore, this cycle attains the power $O(N)$ and the efficiency $\eta_{\mathrm{Car}}-O(1/N)$.
We emphasize that the cycle is not slow-regime and the temperatures $\beta_H$ and $\beta_C$ are arbitrary.

\subsection{Steady current without dissipation in $2N$-state model 1: temperature difference}
We give an example of the steady current with constant-order entropy production rate in the $2N$-state model.
Again, we use the system Hamiltonian
\begin{align}
H=\hbar\omega\sum^{N}_{j=1}\ket{{\rm e},j}\bra{{\rm e},j}.
\end{align}
We use two baths, hot one and cold one, and define the interaction Hamiltonians between each bath and the system as follows
\begin{align}
H^{H}_{\mathrm{int}}=b_H\sum^{N}_{j,j'=1}(\sigma^{j,j'}_{+}\otimes B
+\sigma^{j,j'}_{-}\otimes B^{\dagger}),\enskip\enskip
\mbox{and}\enskip\enskip
H^{C}_{\mathrm{int}}=b_C\sum^{N}_{j,j'=1}(\sigma^{j,j'}_{+}\otimes B
+\sigma^{j,j'}_{-}\otimes B^{\dagger}).
\end{align}
Then, the time evolution of the system is
\begin{align}
\partial_t \rho&=-\frac{i}{\hbar}[H,\rho]+{\cal D}_H[\rho]+{\cal D}_C[\rho]\\
{\cal D}_{H}[\rho]&=\Gamma^{H,N}_{\downarrow}\left[L\rho L^{\dagger}-\frac{1}{2}\{L^{\dagger}L,\rho\}\right]+\Gamma^{H,N}_{\uparrow}\left[L^{\dagger}\rho L-\frac{1}{2}\{LL^{\dagger},\rho\}\right]\\
{\cal D}_{C}[\rho]&=\Gamma^{C,N}_{\downarrow}\left[L\rho L^{\dagger}-\frac{1}{2}\{L^{\dagger}L,\rho\}\right]+\Gamma^{C,N}_{\uparrow}\left[L^{\dagger}\rho L-\frac{1}{2}\{LL^{\dagger},\rho\}\right],
\end{align}
where $L$ is given in Eq.~\eqref{defL}, and $\Gamma^{H,N}_{\uparrow\downarrow}$ and $\Gamma^{C,N}_{\uparrow\downarrow}$ satisifies $\Gamma^{H,N}_{\downarrow}/\Gamma^{H,N}_{\uparrow}=e^{\beta^{(N)}_H\hbar\omega}$ and $\Gamma^{C,N}_{\downarrow}/\Gamma^{C,N}_{\uparrow}=e^{\beta^{(N)}_C\hbar\omega}$.
Here $\beta^{(N)}_{H}$ and $\beta^{(N)}_{C}$ are the temperatures of the hot bath and the cold bath.
Noting that $\Gamma^{H}_{\downarrow}$ and $\Gamma^{C}_{\downarrow}$ are proportional to $b_H^2$ and $b_C^2$, we take $b_C$, $b_H$, $\beta^{(N)}_{H}$ and $\beta^{(N)}_{C}$ satisfying
\begin{align}
(\beta^{(N)}_C-\beta^{(N)}_H)\hbar\omega&=\log\frac{1+\frac{1}{N}}{1-\frac{1}{N}}\label{HCcond},\\
\Gamma^{H,N}_{\downarrow}&=\Gamma^{C,N}_{\downarrow}=\mbox{const}.
\end{align}
Then,
\begin{align}
\frac{(1+\frac{1}{N})\Gamma^{H,N}_{\downarrow}}{\Gamma^{H,N}_{\uparrow}}=\frac{(1-\frac{1}{N})\Gamma^{C,N}_{\downarrow}}{\Gamma^{C,N}_{\uparrow}}
\end{align}
Therefore, we can take $\rho^{(N)}_{\rm ee}$ and $\rho^{(N)}_{\rm gg}$ as
\begin{align}
(1+\frac{1}{N})\Gamma^{H,N}_{\downarrow}\rho^{(N)}_{\rm ee}&=\Gamma^{H,N}_{\uparrow}\rho^{(N)}_{\rm gg}\\
(1-\frac{1}{N})\Gamma^{C,N}_{\downarrow}\rho^{(N)}_{\rm ee}&=\Gamma^{C,N}_{\uparrow}\rho^{(N)}_{\rm gg}.
\end{align}
Then, the state 
$\rho^{(N)}:=\rho^{(N)}_{\rm gg}\ket{{\rm g},+}\bra{{\rm g},+}+\rho^{(N)}_{\rm ee}\ket{{\rm e},+}\bra{{\rm e},+}$ is a steady state, due to the following:
\begin{align}
{\cal D}_{H}[\rho]+{\cal D}_{C}[\rho]=&N^2(\Gamma^{H,N}_{\downarrow}\rho^{(N)}_{\rm ee}-\Gamma^{H,N}_{\uparrow}\rho^{(N)}_{\rm gg}+\Gamma^{C,N}_{\downarrow}\rho^{(N)}_{\rm ee}-\Gamma^{C,N}_{\uparrow}\rho^{(N)}_{\rm gg})\ket{{\rm g},+}\bra{{\rm g},+}\nonumber\\
&-N^2(\Gamma^{H,N}_{\downarrow}\rho^{(N)}_{\rm ee}-\Gamma^{H,N}_{\uparrow}\rho^{(N)}_{\rm gg}+\Gamma^{C,N}_{\downarrow}\rho^{(N)}_{\rm ee}-\Gamma^{C,N}_{\uparrow}\rho^{(N)}_{\rm gg})\ket{{\rm e},+}\bra{{\rm e},+}\nonumber\\
=&N^2(-\frac{1}{N}\Gamma^{H,N}_{\downarrow}+\frac{1}{N}\Gamma^{H,N}_{\downarrow})\rho^{(N)}_{0}\ket{{\rm g},+}\bra{{\rm g},+}
-N^2(-\frac{1}{N}\Gamma^{H,N}_{\downarrow}+\frac{1}{N}\Gamma^{H,N}_{\downarrow})\rho^{(N)}_{0}\ket{{\rm e},+}\bra{{\rm e},+}
\nonumber\\
=&0
\end{align}
The heat current of this state is 
\begin{align}
J^{H}[\rho^{(N)}]&=N^2\hbar\omega(\Gamma^{H,N}_{\uparrow}\rho^{(N)}_{\rm gg}-\Gamma^{H,N}_{\downarrow}\rho^{(N)}_{\rm ee})\nonumber\\
&=N\hbar\omega\Gamma^{H,N}_{\downarrow}\rho^{(N)}_{\rm ee}\nonumber\\
&=N\hbar\omega\Gamma^{C,N}_{\downarrow}\rho^{(N)}_{\rm ee}\nonumber\\
&=-J^{C}[\rho^{(N)}].
\end{align}
The entropy production rate is
\begin{align}
\dot{\sigma}(\rho)&=N^2(\Gamma^{H,N}_{\uparrow}\rho^{(N)}_{\rm gg}-\Gamma^{H,N}_{\downarrow}\rho^{(N)}_{\rm ee})\log\frac{\Gamma^{H,N}_{\uparrow}\rho^{(N)}_{\rm gg}}{\Gamma^{H,N}_{\downarrow}\rho^{(N)}_{\rm ee}}
+
N^2(\Gamma^{C,N}_{\uparrow}\rho^{(N)}_{\rm gg}-\Gamma^{C,N}_{\downarrow}\rho^{(N)}_{\rm ee})\log\frac{\Gamma^{C,N}_{\uparrow}\rho^{(N)}_{\rm gg}}{\Gamma^{C,N}_{\downarrow}\rho^{(N)}_{\rm ee}}\nonumber\\
&=N^2\frac{1}{N}\Gamma^{H,N}_{\downarrow}\rho^{(N)}_{\rm ee}\log (1+\frac{1}{N})
-N^2\frac{1}{N}\Gamma^{C,N}_{\downarrow}\rho^{(N)}_{\rm ee}\log (1-\frac{1}{N})=O(1)
\end{align}
Therefore, we obtain $O(N)$ steady heat current with $O(1)$ entropy production rate.

\subsection{Steady current without dissipation in $2N$-state model 2: chemical potential difference}\label{chemi}

Finally, we give an example of the steady current with constant-order entropy production rate between two heat baths whose chemical potentials are different.
We use the system Hamiltonian
\begin{align}
H=\hbar\omega\sum^{N}_{j=1}\ket{{\rm e},j}\bra{{\rm e},j}
\end{align}
We employ Bosonic bath whose Hamiltonian is $\hbar\omega a^{\dagger}a$.
We use two of the Bosonic baths $B_L$ and $B_R$, whose states are in the following ground canonical states:
\begin{align}
\rho_{L}=\frac{e^{-\beta(\hbar\omega-\mu^{(N)}_{L})a^{\dagger}a}}{\Tr[e^{-\beta(\hbar\omega-\mu^{(N)}_{L})a^{\dagger}a}]},\enskip 
\rho_{R}=\frac{e^{-\beta(\hbar\omega-\mu^{(N)}_{R})a^{\dagger}a}}{\Tr[e^{-\beta(\hbar\omega-\mu^{(N)}_{R})a^{\dagger}a}]}
\end{align}
Here $\mu^{(N)}_{L}$ and $\mu^{(N)}_{R}$ are chemical potentials of the baths.
We define the interaction Hamiltonians between each bath and the system as follows
\begin{align}
H^{L}_{\mathrm{int}}=b_L\sum^{N}_{j,j'=1}(\sigma^{j,j'}_{+}\otimes a
+\sigma^{j,j'}_{-}\otimes a^{\dagger}),\enskip\enskip
\mbox{and}\enskip\enskip
H^{R}_{\mathrm{int}}=b_R\sum^{N}_{j,j'=1}(\sigma^{j,j'}_{+}\otimes a
+\sigma^{j,j'}_{-}\otimes a^{\dagger}).
\end{align}
Then, the time evolution of the system is
\begin{align}
\partial_t \rho&=-\frac{i}{\hbar}[H,\rho]+{\cal D}_L[\rho]+{\cal D}_R[\rho]\\
{\cal D}_{L}[\rho]&=\Gamma^{L,N}_{\downarrow}\left[L\rho L^{\dagger}-\frac{1}{2}\{L^{\dagger}L,\rho\}\right]+\Gamma^{L,N}_{\uparrow}\left[L^{\dagger}\rho L-\frac{1}{2}\{LL^{\dagger},\rho\}\right]\\
{\cal D}_{R}[\rho]&=\Gamma^{R,N}_{\downarrow}\left[L\rho L^{\dagger}-\frac{1}{2}\{L^{\dagger}L,\rho\}\right]+\Gamma^{R,N}_{\uparrow}\left[L^{\dagger}\rho L-\frac{1}{2}\{LL^{\dagger},\rho\}\right]
\end{align}
where $L$ is given in Eq.~\eqref{defL}, and $\Gamma^{L,N}_{\uparrow\downarrow}$ and $\Gamma^{R,N}_{\uparrow\downarrow}$ satisifies $\Gamma^{L,N}_{\downarrow}/\Gamma^{L,N}_{\uparrow}=e^{\beta(\hbar\omega-\mu^{(N)}_{L})}$ and $\Gamma^{R,N}_{\downarrow}/\Gamma^{R,N}_{\uparrow}=e^{\beta(\hbar\omega-\mu^{(N)}_{R})}$.
Noting that $\Gamma^{L}_{\downarrow}$ and $\Gamma^{R}_{\downarrow}$ are proportional to $b_L^2$ and $b_R^2$, we take $b_L$, $b_R$, $\mu^{(N)}_{L}$ and $\mu^{(N)}_{R}$ satisfying
\begin{align}
(\mu^{(N)}_L-\mu^{(N)}_R)\hbar\omega&=\log\frac{1+\frac{1}{N}}{1-\frac{1}{N}}\label{HCcond'},\\
\Gamma^{L,N}_{\downarrow}&=\Gamma^{R,N}_{\downarrow}=\mbox{const}.
\end{align}
Then,
\begin{align}
\frac{(1+\frac{1}{N})\Gamma^{L,N}_{\downarrow}}{\Gamma^{L,N}_{\uparrow}}=\frac{(1-\frac{1}{N})\Gamma^{R,N}_{\downarrow}}{\Gamma^{R,N}_{\uparrow}}
\end{align}
Therefore, we can take $\rho^{(N)}_{\rm ee}$ and $\rho^{(N)}_{\rm gg}$ as
\begin{align}
(1+\frac{1}{N})\Gamma^{L,N}_{\downarrow}\rho^{(N)}_{\rm ee}&=\Gamma^{L,N}_{\uparrow}\rho^{(N)}_{\rm gg}\\
(1-\frac{1}{N})\Gamma^{R,N}_{\downarrow}\rho^{(N)}_{\rm ee}&=\Gamma^{R,N}_{\uparrow}\rho^{(N)}_{\rm gg}.
\end{align}
Then, the state 
$\rho^{(N)}:=\rho^{(N)}_{\rm gg}\ket{{\rm g},+}\bra{{\rm g},+}+\rho^{(N)}_{\rm ee}\ket{{\rm e},+}\bra{{\rm e},+}$ is a steady state, due to the following:
\begin{align}
{\cal D}_{L}[\rho]+{\cal D}_{R}[\rho]=&N^2(\Gamma^{L,N}_{\downarrow}\rho^{(N)}_{\rm ee}-\Gamma^{L,N}_{\uparrow}\rho^{(N)}_{\rm gg}+\Gamma^{R,N}_{\downarrow}\rho^{(N)}_{\rm ee}-\Gamma^{R,N}_{\uparrow}\rho^{(N)}_{\rm gg})\ket{{\rm g},+}\bra{{\rm g},+}\nonumber\\
&-N^2(\Gamma^{L,N}_{\downarrow}\rho^{(N)}_{\rm ee}-\Gamma^{L,N}_{\uparrow}\rho^{(N)}_{\rm gg}+\Gamma^{R,N}_{\downarrow}\rho^{(N)}_{\rm ee}-\Gamma^{R,N}_{\uparrow}\rho^{(N)}_{\rm gg})\ket{{\rm e},+}\bra{{\rm e},+}\nonumber\\
=&N^2(-\frac{1}{N}\Gamma^{L,N}_{\downarrow}+\frac{1}{N}\Gamma^{L,N}_{\downarrow})\rho^{(N)}_{0}\ket{{\rm g},+}\bra{{\rm g},+}
-N^2(-\frac{1}{N}\Gamma^{L,N}_{\downarrow}+\frac{1}{N}\Gamma^{L,N}_{\downarrow})\rho^{(N)}_{0}\ket{{\rm e},+}\bra{{\rm e},+}
\nonumber\\
=&0
\end{align}
The heat current of this state is 
\begin{align}
J^{L}[\rho^{(N)}]&=N^2\hbar\omega(\Gamma^{L,N}_{\uparrow}\rho^{(N)}_{\rm gg}-\Gamma^{L,N}_{\downarrow}\rho^{(N)}_{\rm ee})\nonumber\\
&=N\hbar\omega\Gamma^{L,N}_{\downarrow}\rho^{(N)}_{\rm ee}\nonumber\\
&=N\hbar\omega\Gamma^{R,N}_{\downarrow}\rho^{(N)}_{\rm ee}\nonumber\\
&=-J^{R}[\rho^{(N)}].
\end{align}
The entropy production rate is
\begin{align}
\dot{\sigma}(\rho)&=N^2(\Gamma^{L,N}_{\uparrow}\rho^{(N)}_{\rm gg}-\Gamma^{L,N}_{\downarrow}\rho^{(N)}_{\rm ee})\log\frac{\Gamma^{L,N}_{\uparrow}\rho^{(N)}_{\rm gg}}{\Gamma^{L,N}_{\downarrow}\rho^{(N)}_{\rm ee}}
+
N^2(\Gamma^{R,N}_{\uparrow}\rho^{(N)}_{\rm gg}-\Gamma^{R,N}_{\downarrow}\rho^{(N)}_{\rm ee})\log\frac{\Gamma^{R,N}_{\uparrow}\rho^{(N)}_{\rm gg}}{\Gamma^{R,N}_{\downarrow}\rho^{(N)}_{\rm ee}}\nonumber\\
&=N^2\frac{1}{N}\Gamma^{L,N}_{\downarrow}\rho^{(N)}_{\rm ee}\log (1+\frac{1}{N})
-N^2\frac{1}{N}\Gamma^{R,N}_{\downarrow}\rho^{(N)}_{\rm ee}\log (1-\frac{1}{N})=O(1)
\end{align}
Therefore, we obtain $O(N)$-order steady heat current with $O(1)$-order entropy production rate.


\section{Details of the numerical calculation using the two-qubit superradiant model}\label{SM-numeric}
In this section, we give details of the numerical calculation presented in the main text. The Hamiltonian of the two-qubit system is given by
\begin{align}
H(t)=\hbar\omega(t)\left( |1\rangle\langle 1| + |2\rangle\langle 2| \right) + 2\hbar\omega(t)|3\rangle\langle 3|,
\end{align}
where $|0\rangle$ is the ground state, $|1\rangle$ and $|2\rangle$ are the states where one of the qubit is excited, and $|3\rangle$ is the state with both qubits being excited. The Lindblad master equation takes the following form
\begin{align}
\partial_{t}\rho_{00} &= \Gamma_{\downarrow}\left( \rho_{11}+\rho_{22}+\rho_{12}+\rho_{21}\right) - 2\Gamma_{\uparrow}\rho_{00} \\
\partial_{t}\rho_{11} &= -\frac{1}{2}(\Gamma_{\uparrow}+\Gamma_{\downarrow})\left( 2\rho_{11}+ \rho_{12}+\rho_{21} \right) + \Gamma_{\uparrow}\rho_{00} + \Gamma_{\downarrow}\rho_{33} \\
\partial_{t}\rho_{22} &= -\frac{1}{2}(\Gamma_{\uparrow}+\Gamma_{\downarrow})\left( 2\rho_{22}+\rho_{12}+\rho_{21} \right) + \Gamma_{\uparrow}\rho_{00} + \Gamma_{\downarrow}\rho_{33}\\
\partial_{t}\rho_{33} &= \Gamma_{\uparrow}\left( \rho_{11}+\rho_{22}+ \rho_{12}+\rho_{21} \right) -2 \Gamma_{\downarrow}\rho_{33} \\
\partial_{t}\rho_{12} &= -\frac{1}{2} (\Gamma_{\uparrow}+\Gamma_{\downarrow}) ( \rho_{11}+\rho_{22} + 2\rho_{12} ) + \Gamma_{\uparrow}\rho_{00} + \Gamma_{\downarrow}\rho_{33},
\end{align}
where $\rho_{ij}=\langle i|\rho|j\rangle$, $\Gamma_{\downarrow}=\Gamma_{0}(1+\exp(-\beta\hbar\omega))^{-1}$ and $\Gamma_{\uparrow}=\Gamma_{0}(1+\exp(\beta\hbar\omega))^{-1}$.  By solving the time-evolution equation described above numerically for the heat engine cycle, we obtained Fig.~3 and Fig.~4 in the main text (choosing $\Gamma_{0}=1$). Here, we choose a block-diagonalized initial state $\rho(0)=\rho_{\bd}(0)$ for the numerical simulation, so the density matrix satisfies $\rho(t)=\rho_{\bd}(t)$ for any $t$. 

We note that the heat current reads
\begin{align}
J(\rho) =J(\rho_{\bd})&=\hbar\omega(t)\left[ 2\Gamma_{\uparrow}\rho_{00}-2\Gamma_{\downarrow}\rho_{33} +(\Gamma_{\uparrow}-\Gamma_{\downarrow})(\rho_{11}+\rho_{22}+\rho_{12}+\rho_{21}) \right], \\
J(\rho_{\sd}) &= \hbar\omega(t) \left[ 2\Gamma_{\uparrow}\rho_{00}-2\Gamma_{\downarrow}\rho_{33} +(\Gamma_{\uparrow}-\Gamma_{\downarrow})(\rho_{11}+\rho_{22}) \right],
\end{align}  
and the upper bound on the current-dissipation ratio reads
\begin{align}
A_{\rm cl}&=[\hbar\omega(t)]^{2}\left\{ \left( \Gamma_{\uparrow}+\Gamma_{\downarrow}\right)(\rho_{11}+\rho_{22}) + 2\Gamma_{\uparrow}\rho_{00}+2\Gamma_{\downarrow}\rho_{33} \right\} , \\
A_{\rm qm}&= 2[\hbar\omega(t)]^{2}\left( \Gamma_{\uparrow}+\Gamma_{\downarrow}\right)|\rho_{12}| ,
\end{align}
by noting that 
\begin{align}
X=[\hbar\omega(t)]^{2}\left\{  (\Gamma_{\uparrow}+\Gamma_{\downarrow})\left( |1\rangle\langle 1|+|2\rangle\langle 2|+|1\rangle\langle 2|+|2\rangle\langle 1|\right) +2\Gamma_{\uparrow}|0\rangle\langle 0|+\Gamma_{\downarrow}|3\rangle\langle 3|\right\}.
\end{align}

\section{Coherence effect on power-efficiency trade-off of heat engines}\label{p-e-t}
In this section, we follow Ref.~\cite{SST} and use (\ref{go}) and obtain the power-efficiency trade-off relation as follows. We denote $Q_{\rm H}>0$ as the heat from the hot bath (at inverse temperature $\beta_{\rm H}$) to the system and $Q_{\rm C}>0$ as the heat from the system to the cold bath (at inverse temperature $\beta_{\rm C}$). Then, for a steady cycle, we have $W=Q_{\rm H}-Q_{\rm C}$ and $\sigma=\beta_{\rm C}Q_{\rm C}-\beta_{\rm H}Q_{\rm H}$. The thermodynamic efficiency is given by $\eta=W/Q_{\rm H}=1-Q_{\rm C}/Q_{\rm H}$ and the Carnot efficiency is given by $\eta_{\rm C}=1-\beta_{\rm H}/\beta_{\rm C}$. By integrating both-hand sides of Eq.~(\ref{go}) and using the Cauchy-Schwartz inequality, we have
\beq
Q_{\rm H}+Q_{\rm C} \leq \sqrt{\frac{1}{2}\tau \bar{A} \sigma }. \label{CCineq}
\eeq
Here, we assume $\dot{Q}>0$ ($\dot{Q}<0$) when the system interacts with the hot (cold) bath. Also, $\tau$ is the time required to complete a cycle, and $\bar{A}=\tau^{-1}\int^{\tau}_{0}dt A$. We further use 
\beq
\eta(\eta_{\rm C}-\eta)=\frac{W}{Q_{\rm H}}\left( \frac{Q_{\rm C}}{Q_{\rm H}}-\frac{\beta_{\rm H}}{\beta_{\rm C}}\right) =\frac{W}{Q_{\rm H}^{2}\beta_{\rm C}} (\beta_{\rm C}Q_{\rm C}-\beta_{\rm H}Q_{\rm H}) = \frac{W\sigma}{Q_{\rm H}^{2}\beta_{\rm C}} .
\eeq
We then find that
\beq
 (Q_{\rm H}+Q_{\rm C})^{2} \leq \frac{1}{2}\tau \bar{A} \frac{\eta(\eta_{\rm C}-\eta)Q_{\rm H}^{2}\beta_{\rm C}}{W},
\eeq
which leads to the following trade-off relation between the power $W/\tau$ and efficiency $\eta$:
\beq
\frac{W}{\tau}\frac{2(2-\eta)^{2}}{\beta_{\rm C}\eta(\eta_{\rm C}-\eta)} \leq \bar{A}. \label{PEtradeoff2}
\eeq

\end{widetext}
\end{document}